\documentclass[12pt,preprint]{aastex}

\slugcomment{To be submitted to \apj}

\shorttitle{A-type Stars with Mid-IR Excesses}
\shortauthors{Hales et al.}

\begin{document}


\title{IPHAS A-type Stars with Mid-IR Excesses in Spitzer Surveys}


\author{Antonio S. Hales\altaffilmark{1,2}, Michael J. Barlow\altaffilmark{2}, Janet E. Drew\altaffilmark{3,4}, Yvonne C. Unruh\altaffilmark{4}, Robert Greimel\altaffilmark{5}, Michael J. Irwin\altaffilmark{6} and Eduardo Gonz\'alez-Solares\altaffilmark{6}} 
\altaffiltext{1}{National Radio Astronomy Observatory, 520 Edgemont Road, Charlottesville, Virginia, 22903-2475, United States; ahales@nrao.edu}
\altaffiltext{2}{Department of Physics and Astronomy, University College London, Gower Street, London, WC1E 6BT, United Kingdom; mjb@star.ucl.ac.uk}
\altaffiltext{3}{Centre for Astrophysics Research, University of Hertfordshire, Hatfield, Hertfordshire AL10 9AB, United Kingdom; j.drew@herts.ac.uk}
\altaffiltext{4}{Imperial College of Science, Technology and Medicine, Blackett Laboratory, Exhibition Road, London, SW7 2AZ, United Kingdom; y.unruh@imperial.ac.uk}
\altaffiltext{5}{Isaac Newton Group of Telescopes, Apartado de Correos 321, E-38700, Santa Cruz de la Palma, Tenerife, Spain}
\altaffiltext{6}{Institute of Astronomy, Madingley Road, Cambridge, CB3 0HA, United Kingdom}

\email{ahales@nrao.edu}

\begin{abstract}
We have identified 17 A-type stars in the Galactic Plane that have
mid-IR excesses at 8~$\mu$m. From observed colors in the
(r$^{\prime}$-H$\alpha$)-$(r^{\prime}-i^{\prime})$ plane, we first
identified 23050 early A-type main sequence (MS) star candidates in
the Isaac Newton Photometric H-Alpha Survey (IPHAS) point source
database that are located in {\em Spitzer} GLIMPSE Galactic Plane
fields. Imposing the requirement that they be detected in all seven
2MASS and IRAC bands led to a sample of 2692 candidate A-type stars
with fully sampled 0.6 to 8~$\mu$m SEDs. Optical classification
spectra of 18 of the IPHAS candidate A-type MS stars showed that all
but one could be well fitted using main sequence A-type
templates, with the other being an A-type supergiant. Out of the 2692
A-type candidates 17 (0.6\%) were found to have 8-$\mu$m excesses
above the expected photospheric values. Taking into account non-A-Type
contamination estimates, the 8-$\mu$m excess fraction is adjusted
to $\sim$0.7\%. The distances to these sources range from
$0.7-2.5$~kpc. Only 10 out of the 17 excess stars had been covered by
{\em Spitzer} MIPSGAL survey fields, of which 5 had detectable
excesses at 24~$\mu$m.  For sources with excesses detected in at least
two mid-IR wavelength bands, blackbody fits to the excess SEDs yielded
temperatures ranging from $270$ to $650$~K, and bolometric luminosity
ratios L$_{\rm IR}$/L$_{\star}$ from
$2.2\times10^{-3}-1.9\times10^{-2}$, with a mean value of
$7.9\times10^{-3}$ (these bolometric luminosities are lower limits as
cold dust is not detectable by this survey). Both the presence of
mid-IR excesses and the derived bolometric luminosity ratios are
consistent with many of these systems being in the planet-building
transition phase between the early protoplanetary disk phase and the
later debris disk phase.
\end{abstract}


\section{Introduction}

Multi-wavelength photometry remains the fundamental tool for detecting
circumstellar (CS) dust around pre-main sequence and main sequence
(MS) stars. Almost all MS stars known to be surrounded by dust disks have 
been
discovered from the shapes of their spectral energy distributions
(SEDs), which show excesses with respect to the stellar photospheres
at infrared (IR) and longer wavelengths \citep{aum84}. Since their
discovery, great interest has arisen in these MS debris disk systems
as they are thought to be signposts of planet formation
\citep{zucker01}. While resolved imaging is crucial for studying the
characteristics of individual systems, photometric surveys remain the
key tool for inferring their statistical evolutionary properties
\citep{meyer07}. Many surveys for dusty MS stars have used the {\em
Infrared Astronomical Satellite} ({\em IRAS}) database
\citep{aumann85,walker88,cheng92,oudmaijer92,mann98,syl00,silv00,rhee07}.
Cross-correlation with spectral catalogs allows a search for excess
fluxes when compared to the expected photospheric emission.

The systems can be characterized by the amount of light absorbed and
re-emitted by the disk (the disk-to-star bolometric luminosity ratio
L$_{\rm IR}$/L$_{\star}$).  Large values of L$_{\rm IR}$/L$_{\star}$
($>10^{-2}$) are associated with massive ($M>0.02\,M_{\odot}$)
orbiting CS disks, typically found around young pre-MS systems (t
$\sim10^6$~Myr), such as T Tauri and HAeBe stars. These, in addition,
often show spectroscopic signatures of material still falling
onto the central star \citep{waters98}. On the other hand, low L$_{\rm
IR}$/L$_{\star}<10^{-3}$ values correspond to MS stars surrounded by
older, more tenuous dusty disks (usually gas-depleted), likely to be
sustained by collisions between larger bodies (debris-disks).\\

CS disks can also be characterized by the wavelength at which the excess
first appears. Excesses detected over a broad wavelength range (from
near-infrared to millimeter wavelengths) are indicative of CS disks
having dust belts located at a range of orbital radii
\citep[e.g.,][]{dull07}. Short wavelength excesses
($\lambda<3$~$\mu$m) indicate the presence of hot dust located in the
innermost regions of the disk (r$<1$~AU). Conversely, the lack of
short wavelength excess flux can be modeled as due to disks with cleared
inner regions, that could be caused by the presence of shepherding
planets \citep[e.g.,][]{calvet02,dalessio05}. Among disks with
cleared inner region two categories are distinguishable: young
gas-rich systems transitioning between their pre-MS and MS stages
\citep[e.g. 49 Cet;][]{hughes08}, and older debris systems that lack
gas as well as planetesimal belts in the innermost regions of the
disk.

The {\em IRAS} samples showed that the number of objects with
12-$\mu$m excesses (but no near-IR excesses, e.g. $\beta$ Pictoris)
was significantly lower than those with excesses at longer
wavelengths, $0.5-2\%$ versus $10-20\%$ for excesses at
$\lambda\ge25$~$\mu$m
\citep{aumann91,cheng92,plets99,fajardo00,lagrange00,song01}.  The
ubiquity of cold disks is consistent with evolutionary scenarios in
which disk clearing occurs in an in-out way; material occupying the
inner region of a protoplanetary disk, and its observational
signatures, disappears before material in the outer disk
\citep{backmann93,wuchter00,meyer07,cieza07}.


With notably improved sensitivity compared to its predecessors, {\em
Spitzer Space Telescope} observations have confirmed that not only do
the occurrence and magnitude of infrared excesses decay with time, but
that in the case of MS systems, {\it{cool}} dust excesses
($\lambda>24$~$\mu$m, $T<100$~K) are systematically more frequent than
{\it{hot}} or {\it{warm}} ones (detected between 2-24~$\mu$m). The
results of \citet{beich05}, \citet{bryden06} and \citet{su06} clearly
indicate that 70-$\mu$m excesses are more common than 24-$\mu$m
excesses for MS stars of a given age, but the different incidence
fractions quoted appear to be dominated by the varying volumes and
targeted spectral types of each sample \citep[FGK stars in the first
two cases, and A-type stars in the work of][]{su06}. Systems with
excesses at mid-IR wavelengths (but without near-IR excesses or other
signs of ongoing accretion) are far less common
\citep{silv06,sicilia06,meyer07,uzpen07}. \citet{silv06} carried out
$3.6-8$~$\mu$m observations of 74 young (t $<30$~Myr), Sun-like stars
($0.7\,M_{\odot}<M<1.5\,M_{\odot}$), in order to investigate the
presence of hot (220-1000~K) dust in the inner regions of CS disks and
found evidence for only 5 optically thick disks, which were more
likely to be pre-MS T Tauri stars and not transition- or $\beta$
Pictoris-like systems.  \citet{uzpen07} cross-correlated the {\em MSX}
\citep{egan97} and {\em Spitzer} GLIMPSE \citep{church05} catalogs
with the Tycho~2 Spectral Catalog \citep{wright03} and inferred the
incidence of 8-$\mu$m excesses to be as low as $1-2$~$\%$ for Galactic
MS systems of spectral type B8 or later. A large fraction of the
GLIMPSE catalog is still unexplored, as most cataloged sources lack
spectral classifications.

In this work we exploit several Galactic Plane surveys in order to
search a very large sample of A-type MS stars for 8-$\mu$m excesses.
One of the main aims of this work was to investigate the incidence of
mid-IR excesses around a large unbiased sample of early A-type
stars. This sample covers previously unexplored magnitude and distance
ranges, allowing us to detect the optical, near-IR and mid-IR
photospheric emission from A-type dwarfs out to distances of
$\sim$2.5~kpc. Being abundant, luminous and devoid of circumstellar
free-free emission \citep[which can lead to false excesses as in the
case of B-stars, e.g.][]{clarke05}, A-type dwarfs are excellent
targets to search for CS dust emission. A large fraction (50\%) of the
debris-disks systems known to date orbit A-type dwarf stars
\citep{silv00,rhee07}. Our MS A-type sample, drawn from the INT/WFS
Photometric H-alpha Survey of the Northern Galactic plane
\citep[IPHAS,][]{drew05,idr07}, is selected from their observed
optical $(r^{\prime}-i^{\prime})$ and $(r^{\prime}-H_{\alpha})$ IPHAS
colors. This color selection scheme ensures most selected objects are
near or on the MS, excluding pre-MS (such as Herbig Ae stars) and
post-MS objects with prominent $H_{\alpha}$
emission. Cross-correlating with near- and mid-IR sources from the
2MASS and GLIMPSE surveys enables their dereddened $0.6$~$\mu$m to
$8$~$\mu$m SEDs to be constructed, in order to search for warm
(T~$\sim150 - 800$~K) mid-IR excesses. In addition, early-release post
basic-calibration $24$-$\mu$m images from the {\em Spitzer} MIPSGAL
survey \citep{carey05} were used to search for $24$-$\mu$m
confirmation of the mid-IR excesses.

Section~\ref{data} describes our method of selecting A-type MS stars
from the IPHAS database and their cross-correlation with the {\em Spitzer}
data. In Section~\ref{excess} we identify those sources with detectable
mid-IR excesses and fit black-bodies to their excesses.  The properties of
the sample are discussed in Section~\ref{disc}, with Section~\ref{conc} 
summarising our conclusions. 

\section{Observations and data processing}\label{data}

\subsection{IPHAS}

\subsubsection{IPHAS Data}

The INT/WFS Photometric H-alpha Survey of the Northern Galactic plane
\citep[IPHAS; ][]{drew05,idr07} is a multi-national observing
programme dedicated to surveying the northern galactic plane. The
northern galactic latitude range of $-5^{\rm o}<b<+5^{\rm o}$
represents a total sky area of 1800 square degrees.

 Two broad-band Sloan r$^{\prime}$ and Sloan i$^{\prime}$ filters, in
conjunction with a narrow band H$\alpha$ filter, provide sufficient
photometric information to identify approximate spectral types for
most stars detected by the survey \citep{drew08,sale08}.  The limiting
magnitude of the IPHAS survey is r$^{\prime}\,\sim\,20.5$.  By the end
of data-taking in 2008-9, over 200 million objects are expected to be
cataloged in terms of their positions and r$^{\prime}$, i$^{\prime}$
and H$\alpha$ magnitudes.  \\

IPHAS observations are made using the Wide Field Camera (WFC) at the
2.5-m Isaac Newton Telescope (INT), La Palma, Spain. The WFC, an
imager made of four 4k$\times2$k EEV CCDs arranged in an L shape,
provides a field of view of $34\times34$~arcmins. The pixel size of
0.3333 arcsec is sufficient to provide high quality sampling of the
$\sim$1 arcsecond seeing encountered at the Observatorio del Roque de
los Muchachos. The r$^{\prime}$ filter has a central wavelength of
$6240$~\AA. The H$\alpha$ filter has a FWHM transmission of $95$~\AA ,
centered at $6568$~\AA, towards the red end of the r$^{\prime}$
filter. The i$^{\prime}$ filter has a central wavelength of
$7743$~\AA.  In order to account for the gaps between CCDs on the WFC,
observations for a given field of view are paired with a second
observation offset $5$ arcmin-West and $5$ arcmin-South. These are
usually referred to as the {\it{on}} and {\it{off}} exposures of a given
field. The total number of pointings required to cover the survey area
is 7635, with most sources being imaged at least twice. The exposure
time in the H$\alpha$ filter was $120$s, while for the r$^{\prime}$
filter the exposure time was $30$s and for the r$^{\prime}$ filter it
was $10$s. The saturation limits (r$^{\prime}\sim13.5$~mag) are
discussed by \citet{drew05} and \citet{idr07}.\\

The data are processed at the Cambridge Astronomical Survey Unit
(CASU\footnote{\tt http://apm2.ast.cam.ac.uk/cgi-bin/wfs/dqc.cgi}) as
described by \citet{irwin85} and \citet{idr07}. Sources are classified
following their extraction characteristics, e.g., {\it{0\,=\,noise
like, 1\,=\,non-stellar, -1\,=\,stellar, -2\,=\,probably stellar,
-9\,=\,saturated}}.  Photometric standards observed during the night
are used for flux calibration in each passband. Astrometric solutions
are initially derived based on the known telescope and camera
characteristics, and then refined using the 2MASS catalogue
\citep{idr07}. The distribution of position discrepancies between
IPHAS and 2MASS is roughly Gaussian, with a peak at 0.0 arcsec, and
has a standard deviation of 0.1 arcsec. Hence, the astrometric
precision of IPHAS with respect to 2MASS is assumed to be better than
0.1 arcsec \citep{idr07}.
%

In recent processing \citep{idr07}, the nightly H$\alpha$ zeropoint is
set at a constant offset of 3.14 with respect to the $r'$ zeropoint,
defined by the flux difference between the narrowband and $r'$
transmission profiles convolved with Vega's spectrum \citep{bohlin07}.
This assures that $(r'-H\alpha)$ is brought to zero for unreddened
Vega-like stars. At the time the work was carried out for this paper
an earlier method of H$\alpha$ calibration was in use that meant the
$(r' - H\alpha)$ was not anchored in this way \citep[see,][]{drew05}.
This had implications for the extraction of A-dwarf candidates which
we discuss below.  After processing, a final catalog for a single
pointing can contain from ten to fifty thousand objects (stellar and
non-stellar).

\subsubsection{A-dwarfs in the $(r^{\prime}-i^{\prime})$ versus $(r^{\prime}-H_{\alpha})$  plane}

\begin{figure}
\begin{center}
\includegraphics[scale=0.9,angle=0,clip=true]
{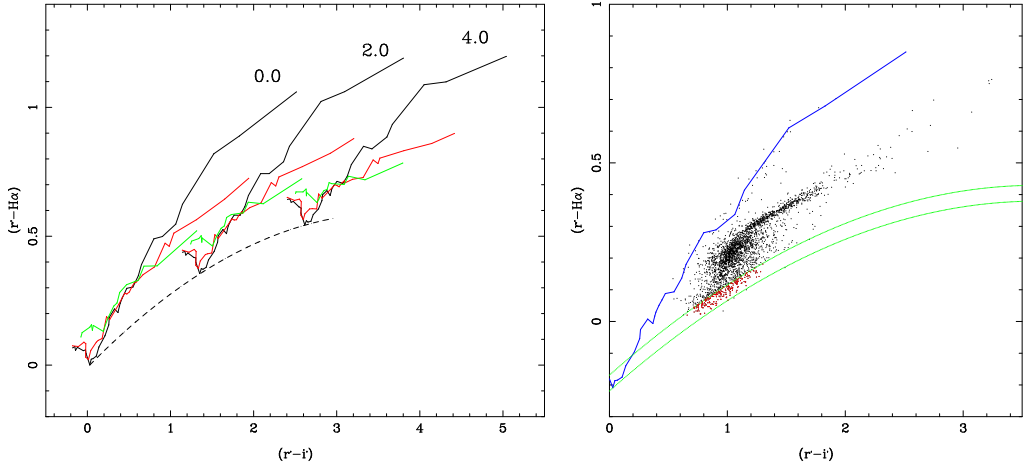}
\caption{{\it{Left-hand panel}}: Synthetic $(r^{\prime}-H_{\alpha})$
versus $(r^{\prime}-i^{\prime})$ colors of normal stars under the
effect of interstellar extinction (adapted from \cite{drew05}). For a
given visual color excess $E(B-V)$, the black lines represent the main
sequence, the red lines represent the giant sequence and the green
lines show the positions of the supergiants. The dashed line shows the
reddening track for the normal metallicity A0V star from the
\citet{pickles98} library: it is a representative early-A reddening
line. {\it{Right-hand panel}}: The $(r^{\prime}-H_{\alpha})$ versus
$(r^{\prime}-i^{\prime})$ color plane for IPHAS field 4232. An
{\it{early-A}} star strip, of width 0.05 mags in
(r$^{\prime}$-H$_{\alpha}$) (green lines) was used to select for
A-type dwarfs (red dots). The blue line corresponds to an unreddened
MS track. }\label{synth}
\end{center}
\end{figure}

The equivalent width of the photospheric H$\alpha$ absorption feature
peaks at early A-types. As a consequence, for a given r$^{\prime}$
magnitude and reddening, the (r$^{\prime}$- H$\alpha$) color of an
A-type star will be a minimum compared to that of other stellar
types. \cite{drew05} computed the synthetic IPHAS
$(r^{\prime}-H_{\alpha})$ and $(r^{\prime}-i^{\prime})$ colors for
stars of various luminosity classes and spectral types present in the
catalog of stellar SEDs by \citet{pickles98}, and investigated their
variations for different amounts of interstellar reddening.  Figure
\ref{synth} (left-panel, adapted from \cite{drew05}), shows a
synthetic IPHAS color-color diagram. The positions of main sequence,
giant and supergiant stars are mapped for three different values of
$E(B-V)$. It can be seen that, as the reddening increases, the minima
of the loci of MS stars trace out an approximately parabolic line. Due
to their strong H$\alpha$ absorption, this line traces the positions
of MS early A-type stars in the $(r^{\prime}-i^{\prime})$,
$(r^{\prime}-H_{\alpha})$ plane as a function of reddening.
Consequently it has been named the {\it{early-A reddening line}}.
Objects near this line will mostly be A0-5 near-MS stars. How this is
so and how these stars may be extracted and exploited has been
presented \citet{drew08}.\\

\subsubsection{Selection of A-dwarf stars}\label{atypefind}

In order to obtain a sample of early A-type stars that can be
cross-correlated with the GLIMPSE point-source catalog, all the
available IPHAS data for regions that overlapped with the GLIMPSE
survey were downloaded ($30^{\rm o}<$ l $<65^{\rm o}$ and $|b|<1^{\rm 
o}$). To ensure the
quality of the sample, we restricted the search to objects having
$r^{\prime}$ magnitudes between $14<m_{r^{\prime}}<18$ and to
fields that had seeing better than $2$~arcsec. In addition,
objects had to be flagged as stellar or probably stellar by CASU in
all 6 IPHAS exposures to be selected ($r^{\prime}$, i$^{\prime}$ and
H$\alpha$ in both on- and off- observations). Only 44\% (134 of 298)
of the IPHAS fields for this region had been obtained at the time that
this investigation commenced. A complete list of the relevant
processed IPHAS catalogs is presented in Table \ref{sumobs}. The
central coordinates of each pointing are listed in both Equatorial
J2000 and Galactic coordinates, arranged in order of increasing
galactic longitude. Column 6 gives the average seeing measured at the
Observatorio del
Roque de los Muchachos during the {\it{on}} and {\it{off}}
integrations for each field. Field numbers (column 1) correspond to
internal IPHAS field names.\\

In each IPHAS field, the lower edge of the main stellar locus in the
$(r^{\prime}-i^{\prime}, r^{\prime}-H_{\alpha})$ plane will always
follow the aforementioned early-A reddening line. Therefore, this line
can be used as the cut line for selecting A-type stars
\citep{drew08}. First, the position of the early-A reddening line had
to be defined interactively, allowing for an empirical
$(r^{\prime}-H_{\alpha})$ shift to deal with the floating $H\alpha$
zero-point magnitude, then in use (in principle the early-A reddening
line should originate close to $(r^{\prime}-i^{\prime}, 
r^{\prime}-H_{\alpha})$ = (0,0), but the variable $H\alpha$
zero-point magnitude requires that the origin of the A-type reddening line
has to be placed interactively for each IPHAS field).

Once a good alignment between the A-type reddening line and the lower
edge of the loci of MS stars has been achieved, all stars located
between the fitted early-A reddening line and a second cutoff line
will be selected as A-type candidates. This defines an early-A
reddening $(r^{\prime}-H_{\alpha})$ strip, starting from the location
of the fitted early A-type reddening line upward (e.g. as in Figure
\ref{synth}, right-hand panel). The choice of width for each field was
made after inspection of the color-color diagram for each field
individually.  The guiding principle was to make this width large
enough to be inclusive of the great majority of A dwarfs.  This
approach inevitably admits some interlopers, mainly of somewhat later
spectral type, but these can be eliminated at the later SED-fitting
stage of the analysis. The width of the $(r^{\prime}-H_{\alpha})$
strip was required to never exceed 0.07~mag and was typically
0.05~mag (cf. the detailed discussion by Drew et al. (2005) of
this point). 
Once the early-A reddening strip had been defined, the
programme identified all stars lying within the strip and extracted
the corresponding stellar coordinates, together with their photometric
information (r$^{\prime}, $i$^{\prime}, $ H$\alpha$ magnitudes,
associated uncertainties, extraction parameters, and 2MASS JHK
photometry, if available).


Table~\ref{sumobs}, column 7, summarizes the numbers of objects
cataloged in each field, which provide the input for A-type star
extractions. Column 8 shows the number of A-type dwarfs extracted for
each IPHAS field. Of a total of
260,223 input stars, 23,050 were selected as early A-type
dwarfs on the basis of their IPHAS colors, corresponding to $8.8\%$ of
the total field stars.

\subsection{GLIMPSE}

GLIMPSE - the Galactic Legacy Infrared Mid-Plane Survey Extraordinaire
\citep{church05} - is a fully sampled, confusion limited, 4-band near-
to mid-infrared survey of the inner third of the Galactic disk, with a
spatial resolution of ~2~arcseconds at the shortest wavelengths
\citep{benjamin03}. Using the Infrared Array Camera
\citep[IRAC,][]{fazio04}, GLIMPSE imaged 220 square degrees at
wavelengths centred on 3.6, 4.5, 5.8, and 8.0~$\mu$m in the Galactic
longitude range 10 to 65 degrees on both sides of the Galactic Centre,
over a Galactic latitude range of +/- 1 degrees. The photometric
sensitivity of 0.3~mJy achieved by GLIMPSE at 8~$\mu$m in lower
background areas (see Table~\ref{iphaszeropoints}) can allow the
unreddened photospheres of early A-type dwarfs to be detected out to
nearly 2~kpc, and is the deepest mid-IR survey of the Galactic plane
carried out to date. GLIMPSE data products come as point-source
catalogs and flux-calibrated mosaic images for each IRAC band,
available at the {\em Spitzer} Science Centre (SSC\footnote{\tt
http://ssc.spitzer.caltech.edu/}) server. The nominal 8~$\mu$m flux-limit 
of the GLIMPSE Highly-Reliable Catalog
of 10~mJy would imply that bare photospheres of A-type stars can be
detected out to distances of $\sim600$~pc.

The GLIMPSE Highly-Reliable Catalog contains point-like sources
whose selection requirements meet a $95\%$ reliability criterion,
determined by the fact that a source must be detected twice in one
band and at least once in an adjacent band. This is called the
{\it{2+1}} criterion. The fluxes in the two bands satisfying the
{\it{2+1}} criterion must be higher than the flux limits listed in
Table~\ref{iphaszeropoints} and are required to have signal-to-noise
ratios (S/N) greater than 5. For the other two bands, the fluxes may
be lower than the survey flux limits, provided they have S/N$>3$.
Therefore, the GLIMPSE catalog can contain sources significantly fainter 
at 8~$\mu$m than the nominal 10~mJy limit. For the 4~s integrations used 
by GLIMPSE, the estimated 8~$\mu$m 5$\sigma$ point-source sensitivity for 
low backgrounds is 0.4~mJy, equivalent to 13th~magnitude
\footnote{\tt http://www.astro.wisc.edu/sirtf/}. The catalog
contains $\sim31$~million sources. \\

The IPHAS A-type sample was correlated with the GLIMPSE catalog,
imposing a maximal radial separation of 1 arcsec. No multiple
associations were found at this matching radius. Column 9 in
Table~\ref{sumobs} summarizes the results from the IPHAS-GLIMPSE
A-type star correlation procedure. Of the 23050 IPHAS-selected A-type
stars, 15312 have reliable detections in at least 2 IRAC bands. Of
these 15312 objects, 11198 have positive detections in all 2MASS
near-IR bands. Source counts in the GLIMPSE catalog drop by a factor
of three from IRAC 3.6 and 4.5~$\mu$m to IRAC 5.8 and 8~$\mu$m
\citep{benjamin05}. This is reflected in the number of correlated
sources detected at 5.8 and 8~$\mu$m (5111 and 2751 stars
respectively, as shown in Table~\ref{gliphas}).  Imposing the
requirement that the selected stars must be detected in all 2MASS and
IRAC bands led to a sub-sample of 2692 A-type stars with fully sampled
$0.6$ to $8$-$\mu$m SEDs. These 2692 stars obtained from the
IPHAS/GLIMPSE correlation will be referred as the GLIPHAS sample
hereafter.  \\

\begin{figure}[h!]
\begin{center}
\includegraphics[scale=0.5,angle=-90,clip=true]
{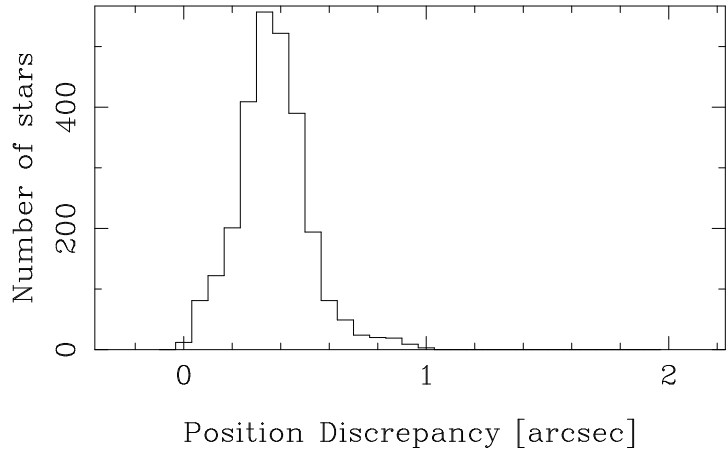}
\caption{ Distributions of radial distances between the IPHAS and
GLIMPSE positions.
\label{histo_deltas}}
\end{center}
\end{figure}

The distribution of positional discrepancies between the GLIMPSE
positions versus the 2MASS positions for the GLIPHAS sample peaks at
0.23 arcsec, with typical dispersions of 0.07 arcsec (as described in
both the GLIMPSE Data Release and in the GLIMPSE Quality Assurance
documentation available in the Team's website). Therefore, there is an
average systematic offset of 0.2-0.3 arcsec between the GLIMPSE and
2MASS positions which should reflect in the IPHAS-GLIMPSE correlation
(since the IPHAS positions are tied to 2MASS).  The distribution of
radial distances between the IPHAS and GLIMPSE positions is shown in
Figure~\ref{histo_deltas}. The distribution is approximately Gaussian,
peaks at 0.36~arcsec and has a standard deviation of 0.11 arcsec. The
mean IPHAS-GLIMPSE offset of 0.36~arcsec is slightly larger than the
0.2-0.3~arcsec expected by assuming the IPHAS-2MASS mean offset is
zero and the GLIMPSE-2MASS mean offset is 0.23~arcsec. However, as
stated in the GLIMPSE Data Release documentation, the offsets between
GLIMPSE and 2MASS sources can be larger for fainter sources due to
poor centroiding. As we are tackling the faint end of the GLIMPSE
dataset, we believe this may be responsible for the small increase
between the GLIMPSE-2MASS and IPHAS-GLIMPSE mean offsets. 

The probability of possible confusing sources can be estimated by
considering the average source density of the two catalogues over the
surveyed area, multiplied by the area of each cross-correlation
($\pi\times 1.0^2$).  There are $\sim$7~million GLIMPSE sources within
the surveyed region ($30^{\rm o}<$ l $<65^{\rm o}$ and $|b|<1^{\rm
o}$). However, most IPHAS A-type stars will have 3.6-$\mu$m magnitudes
brighter than 14 (based on reddening estimates from the $(r' - i')$
colors and assuming A-type colors; Section~\ref{excess} for
details). After removing stars with [3.6]$>14$, there are
$6.05\times10^6$ GLIMPSE stars that could possibly contaminate the
IPHAS cross-correlation (corresponding to a source density of
$6.6\times10^{-3}$arcsec$^{-2}$ over the $\sim$70 deg$^{2}$ surveyed
area). Therefore there is a 2\% chance of false positives in the
search for positional coincidences, implying that amongst the 2692
GLIPHAS A-type stars there could be up to 56 contaminating objects
($6.6\times10^{-3}\times\pi\times1.0^2\times2692= 56$).

\subsection{WHT optical spectra of IPHAS-selected A-type 
stars}\label{spectraltypes}

Optical spectra of 18 IPHAS A-type star candidates, 10 of which are in
the GLIPHAS sample, have been acquired in order to test
the reliability of our photometric spectral-type selection method. The
data were acquired using the Intermediate dispersion Spectrograph
Imaging System (ISIS) on the 4.2-m William Herschel Telescope (WHT)
during the nights of 2006 August 23 and 24, using two different
gratings to simultaneously cover the blue (central wavelength
4249~$\rm{\AA}$) and red (7506~$\rm{\AA}$) regions. Seeing conditions
ranged between 1.2 to 2~arcsec on both nights.

Standard data-reduction procedures were performed using the FIGARO
Starlink\footnote{\tt http://www.starlink.rl.ac.uk/}
application. These included bias-subtraction, flat-fielding,
sky-subtraction, extraction, wavelength calibration and relative flux
calibration.  The red exposures showed strong fringing (typically on
scales of the order of 10 to 20 pixels). This, along with other
pixel-to-pixel variations, was removed during the flat-fielding
process. No attempt was made to correct for the grating efficiency
using the flat-field exposures; it was instead removed during the flux
calibration. The spectral resolution as measured from the FWHM of arc
lines was found to be 4.25~$\rm{\AA}$ in the blue and 3.32~$\rm{\AA}$
in the red. The dispersion solution was estimated to be good to
0.01~$\rm{\AA}$ in the red region, while in the blue it was found to
be accurate to 0.04~$\rm{\AA}$. Flux calibration was performed using
the optimal extraction method \citep{horne1986} using BD+28$^{\rm
o}$4211 \citep{Oke90} as a relative flux calibrator.

Spectral classification was performed by comparing the WHT spectra
with early-type template spectra from the Indo-US Library of Coud\'e
Feed Stellar Spectra \citep[$\sim1~\rm{\AA}$ resolution,][]{valdes04},
after degrading the library spectra to match the spectral resolution
of the data. The observed spectra were dereddened using the visual
extinction estimates from the IPHAS colors (Table~\ref{derived24}, see
Section~\ref{8micres} for details), and then matched by eye to the
closest spectral type using a spectral type grid ranging from B6V to
F5V. The Ca~{\sc ii} K line (3933.66~$\rm{\AA}$) is a crucial spectral
type diagnostic for early-type stars, but can potentially be affected
by interstellar Ca~{\sc ii} absorption over 1-2 kpc sight-lines.  The
Ca~{\sc ii} IR triplet (8498.02, 8542.09, and 8662.14~$\rm{\AA}$) does
not suffer from this problem and due to the lower reddening in this
spectral region the lines are observed with good S/N. We therefore
carried out two different spectral type assignments for each star, one
based on the blue spectrum, in which the Balmer lines and Ca~{\sc ii}
K-line of target and template stars were compared, and one based on
the red spectrum, where the Paschen and Ca~{\sc ii} IR-triplet lines
of target and templates were compared. The comparison between the
template spectrum and the 10 GLIPHAS spectra is shown in
Figure~\ref{wht}.  With the exception of WHT~4 (see below), all 18
stars that were observed were classified as near main sequence A-type.
The blue- and red-based spectral type assignments for the 10 GLIPHAS
stars are presented in Table~\ref{speclass}, together with the
measured equivalent widths of the Ca~{\sc ii} K-line and the
IR-triplet lines. The red-based spectral types range from A0V to A5V,
and are systematically earlier than the blue subtypes, by at least a
sub-type. This is as would be expected if the blue-based
classifications suffer from interstellar Ca~{\sc ii}-K contamination,
leading to an apparently later spectral type. In section~\ref{excess}
the dereddened SEDs of all stars with WHT spectra are shown against a
reference A3V stellar SED. In all cases the SED of the observed star
is confirmed to follow the SED of an A-type star. The optical spectra
and the overall SEDs both demonstrate the reliability of our method of
selecting A-type stars based on their IPHAS colors.


No blue flux was detected from WHT~4, so no blue-based spectral
classification was possible. In addition, its IR-triplet lines were
too deep to be fitted by any A-dwarf spectra. This could be explained
by WHT~4 being a giant or a supergiant, since a lower surface gravity
leads to deeper and narrower lines. A higher luminosity star would be
consistent with the lack of blue flux, since it would be located much
further away and more highly reddened.  Comparison of its spectrum
with those of the giant and supergiant templates from the Indo-US
library showed that it could be well matched to the spectrum of an A5
supergiant (A5Ia). For an M$_{\rm v}$ of --7.4 \citep{schmidtk82} this
would correspond to a spectrophotometric distance of 4.6~kpc, close to
the location of the Scutum-Crux Arm \citep{bronfman92,russeil03} and
much more in keeping with its large reddening (A$_{\rm r^{\prime}}$ =
9.3) than the distance of 100~pc that would be derived assuming the
object is a dwarf (see Table~\ref{whtphotom}).

\subsection {MIPSGAL}\label{24micres}

MIPSGAL \citep{carey05} is a {\em Spitzer} legacy survey that covers
220 square degrees of the inner Galactic plane ($65^{\rm o}<l<10^{\rm
o}$ and $-10^{\rm o}<$l$<-65^{\rm o}$ for $|$b$|<1^{\rm o}$) at 24 and
70~$\mu$m with the Multiband Imaging Photometer for {\em Spitzer}
\citep[MIPS,][]{rieke04}. It has significantly better resolution and
sensitivity than previous infrared surveys covering the plane at these
wavelengths. MIPSGAL complements the Galactic Disk area covered by
GLIMPSE. The MIPSGAL survey was not complete at the time this
investigation was begun, with only post basic-calibration (PBCD)
mosaic images of a few sky regions being available.

In Section~\ref{excess}, we searched for GLIPHAS stars with excess
8~$\mu$m fluxes with respect to the expected photospheric
emission. For the 17 GLIPHAS stars found to have 8 $\mu$m excesses
(presented in Section~\ref{8micres}), we inspected mosaic images from
the MIPSGAL survey in order to search for possible 24-$\mu$m
detections. The availability of 24-$\mu$m data-points can be used as
confirmation of a mid-IR excess and to provide better constraints when
modelling the properties of the emitting dust.

Since only the 24-$\mu$m PBCD images could be used for science
analysis\footnote{see the MIPS data-handbook, at
{\tt{http://ssc.spitzer.caltech.edu/mips/dh/}}}, only the 24-$\mu$m
images were downloaded. The images are calibrated in units of MJy/sr
and so can be used for flux estimation purposes. The zeropoint used by
us to convert 24-$\mu$m fluxes to magnitudes was 7.14~Jy, as computed
from an extrapolation of a model spectrum of Vega (provided by the
SSC). Based on preliminary analysis of mosaic images for
low-background regions, sources with fluxes down to 1 mJy could be
extracted. This corresponds to a 24~$\mu$m magnitude of approximately
9.6.\\

The MIPS images were visually inspected at the positions of each of
the 17 GLIPHAS stars with 8 $\mu$m excesses (presented in
Section~\ref{8micres}, Tables~\ref{24excess} and \ref{derived24}). 5
of the 17 stars had point-like counterparts located within 1 pixel
(2.45~arcseconds) of the registered IPHAS position, and were selected
for flux extraction. Upper-limits for 5 other stars were also derived,
while the positions of the remaining 7 stars had not been covered by
the currently available MIPSGAL images. 5$\sigma$ 24-$\mu$m upper
limits for the 5 non-detected sources were computed using the formula
given by \citet{uzpen07}, assuming a diffraction-limited aperture size
and a rms flux limit computed within a $25\times25$ pixel box centered
at the position of the star.

The 24 $\mu$m fluxes of the detected stars were measured in the
following way: an average Point Response Function (PRF) was
constructed for each image by selecting a few stars of similar
brightness to the candidate star. PRF reference stars located close to
the candidate were preferred in order to avoid known PRF variations
across the mosaic images. The Image Reduction and Analysis Facility
(IRAF\footnote{\tt http://iraf.noao.edu/ }) package DAOPHOT was used
to create the PRF and to extract the candidate object flux.\\

Once created, this PRF was used to fit the image of the candidate star
and to extract its flux, using an aperture of 5 pixels
(12.45~arcseconds). The sky background was defined as the mode value
within an annular region of $4$ pixels in width and located $9$~pixels
in radius away from star. This background estimation was subtracted
when calculating the star's flux. Photometric uncertainties are the
statistical errors. Residual images were produced and inspected for
possible extraction artifacts or over/under-subtraction. An aperture
correction of 1.17 was applied to the derived fluxes, taken from the
MIPS Data Handbook.\\

For all 5 detected sources the fitted positions agreed to within less
than 0.5 arcseconds with the cataloged IPHAS sky coordinates,
suggesting that they are indeed associated with the A-type stars from
our IPHAS-GLIMPSE sample. The 24-$\mu$m fluxes and upper limits are
presented and discussed in Section~\ref{excess}. \\

%

\section {A-type Stars with Mid-infrared Excesses}\label{excess}

\subsection {IPHAS-GLIMPSE A-stars with 8-$\mu$m excesses}\label{8micres}

We searched for stars with 8-$\mu$m excesses amongst the 2692 A-type
stars from the GLIPHAS sample, initially by looking for unusually
large $(K-8)$ colors \citep[e.g.,][]{aumann91,uzpen05,uzpen07}. Since
the stars of our sample should be mostly early A-type stars, their
dereddened colors correspond closely to their color excesses. Our
IPHAS color-color selection criterion is expected to extract stars
with spectral types mainly in the range A0V to A5V. Consistent with
the discussion presented by \citet{drew08}, we adopt $M_{r'} = 1.55$,
and intrinsic $ (r^{\prime}-i^{\prime})= 0.05$ as representative of
the target early-A dwarfs (cf. \citet{houk97} Hipparcos absolute
magnitudes, and \citet{kenyon95} colors).

The observed magnitudes were dereddened using the optical and near-IR
reddening laws of \citet{schlegel98} (after converting the 2MASS
magnitudes to UKIRT $JHK$ magnitudes using the solutions from
\citet{carpenter01}, as the \citet{schlegel98} extinction coefficients
are in the UKIRT system). For the IRAC bands, the extinction
coefficients derived by \citet{indebetouw05} were used.

As outlined by \citet{drew08}, the color excess due to extinction,
E(B-V), can be derived by subtracting the intrinsic color (0.05) from
the observed ($r^{\prime}-i^{\prime}$) color to give
E($r^{\prime}-i^{\prime}$), which is then multiplied by 1.55. To
obtain the visual extinction, A$_v$, E($r^{\prime}-i^{\prime}$) should
be multiplied by 4.901 \citep{schlegel98}.

However, given the long sightlines addressed in this work, it makes
more sense to refer all reddenings to either the r' or the i' band,
rather than to V in view of the fact that atypical laws vary away from
the norm most strongly in the blue-to-visual part of the optical
spectrum \citep{cardelli89}. Specifically, the following relations
were applied: A$_{r^{\prime}}$=4.13 $\cdot$
E($r^{\prime}-i^{\prime}$), and A$_{i^{\prime}}$=3.13$\cdot$
E($r^{\prime}-i^{\prime}$) \citep[using tabulated data
from][]{schlegel98}.

Color excesses were calculated by taking the differences between the
dereddened magnitudes, since the intrinsic colors are expected to be
close to 0.0 for early A-type stars. Figure~\ref{fig:k8} shows the
dereddened $(J - H, K - 8)$ color-color diagram for the 2692 stars
with measured fluxes in all IPHAS, 2MASS and IRAC bands. The
dereddened $(J - H)$ colors of the sample cluster around a mean of
$0.03$, consistent with most of them being A stars \citep[recalling
that mean $(J - H)$ for unreddened A0-5 stars ranges from 0.0 to
0.06;][]{kenyon95}. The standard deviation, $0.12$ mags, will be mainly
due to photometric errors, which are amplified in the dereddening
process, and perhaps partially due to contamination of the sample by
stars that are not A type.  Indeed there is a modest redward skew of
the dereddened $(J - H)$ distribution that could be induced by
contaminant objects, being more commonly of later spectral type than
A0-5, than earlier: specifically 36 stars (1.3~\% of the total sample)
are present with $(J - H) > 0.36$, 3$\sigma$ more red than the mean,
while only 1 star (0.04 \%) lies in the blue tail with $(J - H) <
0.04$.  Furthermore, whilst it is likely that the majority of the
sample are near the main sequence, it is important to bear in mind
also that this color selection on its own does not exclude more
luminous evolved objects.  We can place a figure on the likely level
of contamination after examining the excess $(K - 8)$ objects.

$E(K - 8)$ color excesses were calculated by taking the differences
between the dereddened $K$ and 8-$\mu$m magnitudes, since the
intrinsic $(K - 8)$ color is expected to be close to 0.0 for A0-5
stars.  The distribution of $E(K - 8)$ is not Gaussian, as shown in
Figure~\ref{fig:distributions}, right-hand panel (the Shapiro-Wilk
test for normality was performed and the null hypothesis of normality
was rejected at 99\% confidence level). It has a mean of 0.05, a
standard deviation of 0.29, and a median of 0.00. 34 objects with $E(K
- 8)$ larger than 3 times the standard deviation from the mean
(i.e. $E(K - 8)>0.9$) were removed and statistics recomputed for the
main sample. This procedure was repeated until the statistics
converged to a mean $E(K - 8)=0.03$ and standard deviation of
0.22. The Shapiro-Wilk test was recomputed but again argued against
normality of the sigma-clipped sample.  The width of the $E(K - 8)$
distribution for the sigma-clipped sample cannot be attributed purely
to photometric errors, as the distribution of $(K - 8)$ errors peaks
at the mean value $\sigma^{2}_{phot}=0.15$ and has a dispersion of
$0.07$. From this comparison one can estimate the additional source of
errors introduced to the color excesses from the dereddening
procedure. Assuming
$\sigma^2_{K-[8])}=\sigma^{2}_{phot}+\sigma^2_{dered}$, the above
numbers imply $\sigma_{dered}=0.16$.


Given that the distribution is not Normal (not even after convergence
of the iterative sigma-clipping of a total of 58 outliers with $E(K -
8)$ larger than 3 times the standard deviation from the mean), we
searched for $E(K - 8)$ excess objects by comparing the difference
between the observed color for every source and the mean of the
sigma-clipped sample, divided by the quadratic sum of each's star
photometric uncertainty and the standard deviation of the clipped
sample. In this manner, we define the signal-to-noise (SN) ratio of an
$E(K - 8)$ excess as $SN=\frac{E(K
-8)-0.03}{(\sigma^2_{K-[8]}+0.22^2)}$. The distribution of SN for the
entire sample of 2692 stars is shown in the left-hand panel of
Figure~\ref{fig:distributions}. The distribution is approximately
Gaussian, peaks at 0.0, and has a positive tail due to the presence of
possible $E(K - 8)$ excess sources. There are no sources with SN$<-3$
and conversely, we adopt SN$>3$ as the threshold for considering an
$E(K - 8)$ excess to be present. A total of 20 stars are found to have
SN$>3$, and are highlighted by diamond symbols in the color-color
diagram in Figure~\ref{fig:k8}.

Assuming that the K-band flux is photospheric and that the color
excess is due purely to an 8 micron flux excess, one could conclude
these 20 stars with 8-$\mu$m excesses correspond to a real population
of objects with continuum excesses. The SEDs of the twenty 8-$\mu$m
excess stars have been constructed and inspected to check whether they
conform to those of A0-5 stars.  In this step we found that three had
SEDs deviating significantly from this expectation, in that they more
strongly resembled cooler spectral types.  Since the $E(K - 8)$
selection should not bias in favour of an A spectral type, this
failure rate provides us with a rough estimate of the non-A-type
contamination of the IPHAS-selected sample as a whole: 3 interlopers
out of 20 implies $15\pm4 \%$ of the full sample may not be A stars.
All three stars are excluded from further analysis.  Follow-up
spectroscopy of a subsample of 4 of the 17 excess stars that passed
the SED test confirms that all give a good match to main sequence
A-dwarf template spectra (J191100+094543, J192914+184004,
J192933+183415 and J194541+243253, as listed in
Table~\ref{speclass}). Comparison of the SEDs of the
other 14 IPHAS candidate A-stars having follow-up WHT spectra (and that do 
not have an 8-$\mu$m excess) also indicate they match well an A3V SED, as 
shown in Figure~\ref{fig:sedwht} for the six that are in the GLIPHAS 
sample.


Table~\ref{whtphotom} lists the optical to mid-IR photometry for the 6
GLIPHAS stars with no distinguishable IR-excess for which WHT spectra 
were obtained, while
Table~\ref{24excess} lists the photometry for the 17 GLIPHAS A-type
stars selected as having 8-$\mu$m excesses. In the 5 cases where a
24-$\mu$m flux was detected and extracted according to the procedure
described in Section~\ref{24micres}, the recovered 24$\mu$m
magnitudes are also listed. The 24-$\mu$m fluxes were not dereddened,
as extinction is expected to be negligible at this wavelength.

\subsection {Modeling the mid-IR excesses}\label{8micmod}

In order to characterise the strength of the mid-IR excesses, we
 employed a solar-metallicity Kurucz model atmosphere \citep{kurucz93}
 with $T_{eff} = 9000$~K, $\log g = 4.0$ to extrapolate the observed
 $K$-band flux to longer wavelengths. On normalising the model SED
 flux to the observed fluxes in the $K$ band, we obtain a good fit to
 the optical and near-IR datapoints, confirming the reliability of the
 IPHAS A-type selection method. The mid-IR excesses (excess above
 photosphere, in mJy) were calculated by taking the difference between
 the observed flux and the model SED flux (the latter corresponding
 to the in-band stellar flux, computed by convolving the model SED
 with the 2MASS, IRAC and MIPS~24 filter responses respectively,
 following the description outlined in \citet{robitaille07}). In this
 case, the signal-to-noise ratio of the excesses is defined as the
 ratio of the excess (in mJy) to the photometric uncertainty (in mJy)
 of the overall flux measurement, $(F_{\rm IRAC}-F_{\rm
 phot})/\sigma_{\rm IRAC}$ \citep[e.g,][]{rhee07}. $\sigma_{\rm IRAC}$
 includes both the photometric uncertainty listed in the GLIMPSE
 catalog and an absolute calibration uncertainty of 5$\%$ for all IRAC
 bands. For the 24~$\mu$m points we have assumed a $\sigma_{\rm MIPS}$
 of $10\%$ - a rather conservative value given the $4\%$ absolute
 calibration uncertainty listed in the MIPS Data Handbook Version
 3.3.1 but similar to the value used by \citet{uzpen07}.\\

Table~\ref{derived24} lists the derived infrared excess fluxes above
the reference stellar photosphere for the seventeen 8~$\mu$m excess
stars, along with other quantities such as the r$^{\prime}$
extinction, A$_{\rm r^{\prime}}$, and spectrophotometric distances $d$.
These distances were derived from the observed
r$^{\prime}$ magnitudes, corrected for the r$^{\prime}$ band
extinction, A$_{r^{\prime}}$, assuming that the absolute r$^{\prime}$
magnitude is M$_{r^{\prime}}=1.55$ \citep{houk97}. 


The choice of a representative A-type spectral type has a significant
effect on the derived distance and extinction, while it has little
effect on the derived excess flux and corresponding SNR (less than
$1\%$ change in the derived excess flux when assuming spectral types
ranging from A0V to A5V spectral). In order to account for this
effect, A$_{\rm r^{\prime}}$ and $d$ for each star were calculated
assuming all possible spectral types (A0-5) and the rms value added
quadratically to the 1-$\sigma$ formal uncertainties to give the
errors listed in Table~\ref{24excess}.  The mid-IR excess fluxes
determined for the 17 stars are listed in Table~\ref{derived24}, where
only excesses detected with a S/N higher than 3.0 are shown,
consistent with the selection criteria applied over the E[K-8]
distribution.\\

The observed SEDs are assumed to be the sum of the model stellar
atmosphere SED (SED$_{\star}$) and a cooler black body of a given
temperature (SED$_{\rm \small disk }$),
i.e. $SED_{tot}=SED_{\star}+SED_{\rm \small disk}$. For each of the 17
stars in Table~\ref{derived24}, we searched for the model that
minimized the chi-squared ($\chi^{2}$) difference between the modelled
SED$_{tot}$ and the observed data-points.  The search was performed
using the variable-metric routine Migrad of the Minuit package from
CERN\footnote{\tt
http://seal.web.cern.ch/seal/work-packages/mathlibs/minuit/home.html}. The
iterative process finds simultaneously the best fit temperature of the
black body and its angular diameter in the sky (corresponding to the
square-root of the flux normalisation constant). Errors on individual
parameters are estimated by searching the parameter space for the
$\delta\chi^{2}$ = 1 contour.


Table~\ref{derived24} lists the best-fit black body temperature
derived for each of the 17 stars. Figure~\ref{fig:sed1} shows the
results from our black-body fitting routine for the 12 stars with an
8-$\mu$m excess but lacking 24-$\mu$m detections. The SEDs of the 5
stars with excesses at both 8~$\mu$m and 24~$\mu$m are shown in
Figure~\ref{fig:sed2}. The dereddened optical IPHAS, 2MASS and GLIMPSE
data points are plotted in blue, with photometric error-bars. The
solid line in red corresponds to the reference photospheric SED
normalised to the K-band flux. The pink dotted line represents our
best-fit black-body and the solid green line represents the resulting
best-fitting $SED_{tot}$. The optical data points were excluded from
the fitting routine. For the stars lacking 24-$\mu$m detections, the
derived color temperatures provide upper limits to the maximum
temperatures of the disks, as the lack of longer wavelength
measurements provides no constraints on the presence of cooler
material (allowing for the possibility that the parent planetesimal
belts are extended over a range of radii).  Nonetheless, the derived
color temperatures can be used to estimate the fractional bolometric
infrared fluxes due to the warm components, $L_{\rm
IR}/L_{\star}=(A_{disk}/A_{\star})\cdot(T_{disk}/T_{\star})^4$, where
$A_{disk}$ and $A_{\star}$ represent the K-band normalisation factors
and $T_{\star}$=9000~K. These correspond to distance-independent,
best-fit values of L$_{\rm IR}$/L$_{\star}$, with uncertainties
dependent only on the derivation of $T_{disk}$ and $A_{disk}$. For
stars where an excess is detected at more than one wavelength, the
resulting L$_{\rm IR}$/L$_{\star}$ values are presented in the last
column of Table~\ref{derived24}.

\section {Discussion}\label{disc}

We have conducted a search for A-type dwarfs, selected from the IPHAS
survey, that have mid-IR excesses in the GLIMPSE and/or MIPSGAL
surveys. A sample of 2692 A-type stars was extracted by
cross-correlating the optical IPHAS photometry with the mid-IR {\em
Spitzer} GLIMPSE photometry.  Follow-up optical spectroscopic
observations of 18 IPHAS candidate A-type stars confirmed that all
were of A-type, with only one object (6\%) not fitting a main
sequence star template. As expected from the IPHAS color selection
criteria none of the spectra showed emission-line signatures,
confirming that the selected A-stars are likely to be well-established MS
stars. Figure~\ref{histomagd}, left-panel, shows the magnitude
distribution of the sample at both r$^{\prime}$ and 8~$\mu$m, together
with the dereddened magnitudes. After extinction corrections, both
distributions overlap as expected if, as assumed, they are A-type
stars. The right-panel of Figure~\ref{histomagd}, shows the derived
distance distribution of the sample. Our GLIPHAS sample traces
previously unexplored photometric ranges for debris disk systems,
allowing one to study the incidence of mid-IR excesses at distances of
0.5--2~kpc in the Galactic Plane.  {\em IRAS} searches for debris
disks around MS A-type stars were restricted to $<$120~pc
\citep{rhee07}, while {\em MSX} detection limits allowed A-type stars
out to $200-1000$~pc to be surveyed \citep{clarke05,uzpen07}.

For the GLIPHAS sample of 2692 IPHAS-selected candidate A stars, 17
(0.6\%) were found to have 8-$\mu$m excesses with $S/N > 3$. Taking
into account the probable non-A-Type contamination level of $(15 \pm
4)$\% present in the whole sample, the excess fraction needs to be
adjusted upward to $\sim$0.7\%.  The above fraction of stars showing
warm dust excesses is notably smaller than the $\sim15\%$ quoted by
some pre-{\em Spitzer} surveys for dust excesses around MS stars
\citep{plets99,lagrange00} and the 13\% occurrence fraction quoted by
\cite{song01} for A-type stars with ages ranging from 50 Myr to 1
Gyr. These works, however, quantified the excess fractions at
wavelengths beyond 24~$\mu$m and therefore cannot be used for
comparison. On the other hand, the dust excess fraction of 0.7\%
found by us at 8~$\mu$m is similar to the fraction with dust excesses
found at 12~$\mu$m from {\em IRAS} searches, and coincides with {\em
  MSX} and {\em GLIMPSE} Galactic Plane results of \cite{uzpen07}, who
found that 4 out of 391 A-type stars in their sample (1.0\%) showed
8-$\mu$m excesses. Further, our derived fraction of warm excesses
0.7\% is also consistent with the 1.2\% frequency of warm excesses
found in young clusters \citep{hernandez06,currie08,uzpen08}, and with
recent results from \citet{uzpen08} that report a 0.3\% incidence
fraction of warm excesses among 338 field stars.

We find that none of the 8-$\mu$m excess sources show JHK excesses,
which rules out the presence of very hot dust in the inner parts of
their disks. Four objects (24\%) show excesses at wavelengths shorter than
5.8~$\mu$m, imposing strong constraints on the presence of hot
dust. This supports the hypothesis that the objects in this sample are
relatively evolved systems, older than T Tauri and Herbig Ae stars (as
expected from the lack of H$\alpha$ emission, given their
$(r^{\prime}-H_{\alpha})$ colors).  Table~\ref{derived24} lists and
Figure~\ref{histos} displays the blackbody dust temperatures and
L$_{\rm IR}$/L$_{\rm \star}$ values for both the 8~$\mu$m and
24~$\mu$m excess samples.  Only one object has $T_{disk}>500$~K within
the errors. We find that the SEDs of the mid-IR excess sources can be
fitted by blackbodies with temperatures ranging from $270-650$~K (for
those with detected excesses at more than one wavelength) comparable
to recent samples of warm (T$>200$~K) excess candidates identified
with {\em Spitzer} \citep{uzpen05,uzpen07,hernandez06, currie08}.

The derived color temperatures provide only upper limits to
the maximum temperatures of the disks, as the lack of longer
wavelength measurements provides no constraints on the presence of
cooler material (allowing for the possibility that the parent
planetesimal belts are extended over a range of radii).  Based on
simple radiative equilibrium \citep[e.g.][]{currie08}, the derived
disk temperatures would imply parent planetesimal belts confined to
disk regions of $\sim0.8-10$~AU.

The mid-IR excess systems are found to have fractional disk-to-star
luminosity ratios, L$_{\rm IR}$/L$_{\star}$, ranging from
$2.2\times10^{-3} - 1.9\times10^{-2}$, with a mean of
$7.9\times10^{-3}$. But since cooler dust emitting at far-IR
wavelengths may also be present around the above L$_{\rm
IR}$/L$_{\star}$ values are strictly only lower limits. 
However, for systems with detectable mid-IR excesses for which longer 
wavelength data are also available (e.g. the sample of \citet{syl96}), 
most of the excess luminosity appears at the shorter wavelengths.
The range of fractional luminosities found here are in between the values 
expected for luminous debris-disk and evolved T Tauri and Herbig Ae/Be 
systems with cleared inner regions.



The low incidence rate of mid-IR excesses found in our survey is
consistent with previous inferences that the inner regions of CS disks
are cleared faster than the outer regions, resulting in a greater
persistence of long wavelength excesses
\citep{hayashi85,backmann93,meyer07,cieza07}.  Most prior examples of
stars with 8-12~$\mu$m excesses had H$\alpha$ emission as well as
near-IR excesses, e.g. as found by \citet{dun97b} for the
\citet{syl96} sample of A-stars with mid-IR excesses; our stars, with
no net H$\alpha$ emission and no noticeable near-IR excesses, are
likely to be only a little older than those stars with H$\alpha$
emission that show 8-12~$\mu$m excesses. Our IPHAS color-selection
method ensures the great preponderance of selected stars will be older
than 10~Myr \citep[as described by][]{drew08}, whilst stars in the age
range 5-10~Myrs are admitted as their H$\alpha$ emission becomes
insignificant. The very small proportion (0.7\%) of near-MS A stars
found to exhibit disk emission at 8~$\mu$m are most likely at the
younger end of the entire sample of A stars that have debris disks. 
\citet{currie08} found that the
fraction of true debris-disks ( L$_{\rm IR}$/L$_{\star}\ll10^{-3}$)
peaks at 10-15~Myr, consistent with the low fraction of warm, $\beta$
Pictoris-like, debris-disks identified in this work.

The identification and detailed study of systems with intermediate
values of L$_{\rm IR}$/L$_{\rm \star}$ (10$^{-3}$ - 10$^{-2}$) is
crucial in order to fully understand the disk-clearing and planet
formation processes. Here we have exploited the current generation of
photometric Galactic surveys to increase substantially the number of
known mid-IR excess A-type MS systems, which are likely to be at this
evolutionary stage. High-resolution spectroscopic follow-up
observations of the sample should be carried out in order to refine
the stellar properties (surface gravity, rotational velocities and
metallicities), allowing them to be placed within the context of other
known samples of debris disks.

\section {Conclusions}\label{conc}

Using results from the IPHAS, GLIMPSE and MIPSGAL Galactic Plane
surveys, we have identified 17 new main sequence A-type systems with
warm excesses at 8~$\mu$m and/or 24~$\mu$m. Optical classification
spectra obtained of 10 of the systems confirmed that all but one were
main sequence A-type stars, the exception being an A5Ia supergiant.
The systems have bolometric excess ratios comparable to those of warm
debris-disk systems, similar to $\beta$ Pictoris.  The overall
fraction of sources with 8~$\mu$m excesses was found to be 0.7~$\%$.
The identification of these new CS disk systems shows the potential of
new surveys to increase substantially our knowledge of the occurrence
and characteristics of CS disks in transition between their primordial
and debris-disk phases.  When both the IPHAS and MIPSGAL surveys are
complete, an analysis of the full data-sets is planned in order to
provide a more complete sample of Galactic Plane main sequence A-type
debris disk stars.

\section*{Acknowledgments}

This work was based partially on observations made with the Isaac
Newton Telescope and the William Herschel Telescopes, which are
operated by the Isaac Newton Group in the Observatorio del Roque de
los Muchachos of the Instituto de Astrof\'{i}sica de Canarias, La
Palma, Spain. IPHAS observing time was made available by the time
allocation committees of the UK, Spain and The Netherlands. The WHT
ISIS spectra were obtained as part of the 2006/7 International Time
Programme: `An IPHAS-based exploration of stellar populations in the
northern Milky Way'.  This work made use of data products from the
GLIMPSE survey, which is a legacy science program of the {\em Spitzer
Space Telescope}, funded by the National Aeronautics and Space
Administration; of the SIMBAD database and other facilities operated
at CDS, Strasbourg, France; and of the 2MASS point-source catalog
available at the NASA/IPAC Infrared Science Archive, which is operated
by the Jet Propulsion Laboratory, California Institute of Technology,
under contract with the National Aeronautics and Space Administration.
ASH carried out part of this work whilst being funded by the PPARC
Gemini - Fundaci\'{o}n Andes UK/Chile studentship programme. Parts of
the analysis presented here made use of the Perl Data Language (PDL),
which can be obtained from www.perl.org.

\clearpage

\begin{deluxetable}{ccccccccc}
\tabletypesize{\tiny}

\tablecaption{Summary of processed IPHAS object catalogs\label{sumobs}}
\tablewidth{0pt}
\tablehead{\colhead{Field ID}   &    \colhead{l}   &       \colhead{b}        &   \colhead{RA}        &       \colhead{DEC}           & \colhead{Seeing} & \colhead{Total  Stars}  &     \colhead{Total A-type}   & \colhead{Total with}\\   
\colhead{}           &\colhead{[deg]}   &   \colhead{[deg]}        &   \colhead{}          &       \colhead{}              & \colhead{[arcsec]}& \colhead{}             &     \colhead{dwarfs}   & \colhead{GLIMPSE Correlations }   
}                
\startdata
   4205 	&  29.58 	&  -0.80 	 & 18:48:8.94   	& -03:21:0.0 	&  0.9 	&  874 	        & 138 	&  0 
\\ 4188 	&  29.73 	&  -0.30 	 & 18:46:40.25  	& -02:59:0.0 	&  1.5 	&  0 	        & 179 	&  3 
\\ 4213 	&  30.15 	&  -0.71 	 & 18:48:53.05  	& -02:48:0.0 	&  1.1 	&  612     	& 50 	&  36 
\\ 4196 	&  30.31 	&  -0.21 	 & 18:47:24.49  	& -02:26:0.0 	&  1.1 	&  942     	& 175 	&  158 
\\ 4222 	&  30.72 	&  -0.62 	 & 18:49:37.12  	& -02:15:0.0 	&  1.0 	&  589     	& 67 	&  60 
\\ 4189 	&  31.04 	&  0.37 	 & 18:46:40.62  	& -01:31:0.0 	&  1.5 	&  1698 	& 136 	&  125 
\\ 4254 	&  31.14 	&  -1.03 	 & 18:51:49.68  	& -02:04:0.0 	&  1.3 	&  1425 	& 175 	&  85 
\\ 4232 	&  31.30 	&  -0.53 	 & 18:50:21.29  	& -01:42:0.0 	&  1.0 	&  2737 	& 214 	&  182 
\\ 4214 	&  31.45 	&  -0.04 	 & 18:48:53.12  	& -01:20:0.0 	&  1.0 	&  2548 	& 359 	&  292 
\\ 4197 	&  31.61 	&  0.45 	 & 18:47:25.01  	& -00:58:0.0 	&  1.0 	&  2043 	& 191 	&  63 
\\ 4265 	&  31.71 	&  -0.94 	 & 18:52:33.68  	& -01:31:0.0 	&  1.5 	&  4327 	& 211 	&  148 
\\ 4242 	&  31.87 	&  -0.45 	 & 18:51:05.41          & -01:09:0.0 	&  1.3 	&  4059 	& 388 	&  265 
\\ 4223 	&  32.03 	&  0.05 	 & 18:49:37.33  	& -00:47:0.0 	&  1.3 	&  3246 	& 423 	&  0 
\\ 4206 	&  32.19 	&  0.54 	 & 18:48:09.30   	& -00:25:0.0 	&  1.0 	&  1183 	& 97 	&  1 
\\ 4275 	&  32.28 	&  -0.85 	 & 18:53:17.65  	& -00:58:0.0 	&  1.3 	&  5864 	& 351 	&  104 
\\ 4190 	&  32.35 	&  1.03 	 & 18:46:41.45  	& -00:03:0.0 	&  1.3 	&  1570 	& 125 	&  6 
\\ 4253 	&  32.44 	&  -0.36 	 & 18:51:49.47  	& -00:36:0.0 	&  1.3 	&  4314 	& 538 	&  0 
\\ 4233 	&  32.60 	&  0.13 	 & 18:50:21.48  	& -00:14:0.0 	&  1.1 	&  2286 	& 306 	&  18 
\\ 4215 	&  32.76 	&  0.63 	 & 18:48:53.54  	&  00:08:0.0 	&  1.1 	&  1164 	& 103 	&  90 
\\ 4285 	&  32.86 	&  -0.77 	 & 18:54:01.56   	& -00:25:0.0 	&  1.5 	&  7301 	& 501 	&  0 
\\ 4263 	&  33.02 	&  -0.27 	 & 18:52:33.49  	& -00:03:0.0 	&  1.4 	&  3737 	& 389 	&  111 
\\ 4243 	&  33.17 	&  0.22 	 & 18:51:05.58    	&  00:19:0.0 	&  1.3 	&  1170 	& 131 	&  112 
\\ 4297 	&  33.43 	&  -0.68 	 & 18:54:45.46  	&  00:08:0.0 	&  1.3 	&  4000 	& 541 	&  377 
\\ 4274 	&  33.59 	&  -0.18 	 & 18:53:17.57  	&  00:30:0.0 	&  1.3 	&  1058 	& 79 	&  77 
\\ 4255 	&  33.75 	&  0.31 	 & 18:51:49.73  	&  00:52:0.0 	&  1.3 	&  885 	        & 62 	&  53 
\\ 4235 	&  33.91 	&  0.80 	 & 18:50:22.03  	&  01:14:0.0 	&  1.2 	&  596 	        & 55 	&  43 
\\ 4309 	&  34.00 	&  -0.59 	 & 18:55:29.42  	&  00:41:0.0 	&  1.1 	&  935 	        & 75 	&  66 
\\ 4286 	&  34.16 	&  -0.10 	 & 18:54:01.61   	&  01:03:0.0 	&  1.4 	&  637 	        & 63 	&  58 
\\ 4266 	&  34.32 	&  0.40 	 & 18:52:33.84  	&  01:25:0.0 	&  1.5 	&  837 	        & 82 	&  67 
\\ 4348 	&  34.42 	&  -1.00 	 & 18:57:41.24  	&  00:52:0.0 	&  1.3 	&  343 	        & 17 	&  14 
\\ 4246 	&  34.48 	&  0.89 	 & 18:51:06.18    	&  01:47:0.0 	&  1.2 	&  549 	        & 34 	&  33 
\\ 4322 	&  34.58 	&  -0.50 	 & 18:56:13.36  	&  01:14:0.0 	&  1.2 	&  835 	        & 55 	&  48 
\\ 4298 	&  34.73 	&  -0.01 	 & 18:54:45.60  	&  01:36:0.0 	&  1.3 	&  961 	        & 94 	&  86 
\\ 4276 	&  34.89 	&  0.48 	 & 18:53:17.88  	&  01:58:0.0 	&  1.0 	&  464 	        & 31 	&  28 
\\ 4361 	&  34.99 	&  -0.91 	 & 18:58:25.07  	&  01:25:0.0 	&  1.3 	&  903 	        & 27 	&  17 
\\ 4256 	&  35.05 	&  0.97 	 & 18:51:50.27  	&  02:20:0.0 	&  1.2 	&  370 	        & 40 	&  26 
\\ 4334 	&  35.15 	&  -0.41 	 & 18:56:57.25  	&  01:47:0.0 	&  1.3 	&  779 	        & 41 	&  30 
\\ 4288 	&  35.47 	&  0.57 	 & 18:54:01.88          &  02:31:0.0 	&  1.5 	&  451 	        & 50 	&  49 
\\ 4299 	&  36.04 	&  0.66 	 & 18:54:45.84  	&  03:04:0.0 	&  1.6 	&  717 	        & 85 	&  81 
\\ 4336 	&  36.45 	&  0.26 	 & 18:56:57.35  	&  03:15:0.0 	&  1.3 	&  313 	        & 28 	&  28 
\\ 4401 	&  36.71 	&  -0.64 	 & 19:00:36.45   	&  03:04:0.0 	&  1.3 	&  706 	        & 72 	&  69 
\\ 4325 	&  37.19 	&  0.84 	 & 18:56:13.71  	&  04:10:0.0 	&  1.5 	&  210 	        & 14 	&  14 
\\ 4415 	&  37.28 	&  -0.55 	 & 19:01:20.27   	&  03:37:0.0 	&  1.6 	&  362 	        & 44 	&  43 
\\ 4388 	&  37.44 	&  -0.06 	 & 18:59:52.68  	&  03:59:0.0 	&  1.5 	&  333 	        & 34 	&  34 
\\ 4402 	&  38.01 	&  0.03 	 & 19:00:36.50    	&  04:32:0.0 	&  1.5 	&  278 	        & 16 	&  15 
\\ 4417 	&  38.58 	&  0.12 	 & 19:01:20.30    	&  05:05:0.0 	&  1.5 	&  439 	        & 24 	&  24 
\\ 4389 	&  38.74 	&  0.61 	 & 18:59:52.75  	&  05:27:0.0 	&  1.5 	&  325 	        & 24 	&  23 
\\ 4459 	&  39.00 	&  -0.28 	 & 19:03:31.64          &  05:16:0.0 	&  1.6 	&  679 	        & 76 	&  74 
\\ 4403 	&  39.32 	&  0.70 	 & 19:00:36.52          &  06:00:0.0 	&  1.4 	&  565 	        & 36 	&  33 
\\ 4504 	&  39.41 	&  -0.68 	 & 19:05:42.97          &  05:27:0.0 	&  1.4 	&  1073 	& 102 	&  99 
\\ 4473 	&  39.57 	&  -0.19 	 & 19:04:15.41          &  05:49:0.0 	&  1.4 	&  647     	& 58 	&  57 
\\ 4416 	&  39.89 	&  0.79 	 & 19:01:20.28          &  06:33:0.0 	&  1.5 	&  817 	        & 50 	&  40 
\\ 4488 	&  40.14 	&  -0.10 	 & 19:04:59.19          &  06:22:0.0 	&  1.5 	&  1008 	& 129 	&  92 
\\ 4458 	&  40.30 	&  0.39 	 & 19:03:31.62          &  06:44:0.0 	&  1.7 	&  1310 	& 154 	&  147 
\\ 4534 	&  40.55 	&  -0.50 	 & 19:07:10.52          &  06:33:0.0 	&  1.4 	&  1059 	& 97 	&  91 
\\ 4503 	&  40.71 	&  -0.01 	 & 19:05:42.95    	&  06:55:0.0 	&  1.1 	&  1176 	& 104 	&  100 
\\ 4518 	&  41.28 	&  0.09 	 & 19:06:26.71          &  07:28:0.0 	&  1.1 	&  1213 	& 82 	&  80 
\\ 4593 	&  41.54 	&  -0.80 	 & 19:10:05.70   	&  07:17:0.0 	&  1.1 	&  903   	& 73 	&  72 
\\ 4533 	&  41.85 	&  0.18 	 & 19:07:10.48   	&  08:01:0.0 	&  1.6 	&  986  	& 78 	&  70 
\\ 4502 	&  42.01 	&  0.67 	 & 19:05:42.78    	&  08:23:0.0 	&  1.2 	&  1010 	& 111 	&  109 
\\ 4609 	&  42.11 	&  -0.71 	 & 19:10:49.52  	&  07:50:0.0 	&  1.5 	&  773  	& 46 	&  30 
\\ 4517 	&  42.59 	&  0.76 	 & 19:06:26.46 	        &  08:56:0.0 	&  1.2 	&  1005 	& 80 	&  78 
\\ 4624 	&  42.68 	&  -0.61 	 & 19:11:33.34  	&  08:23:0.0 	&  1.5 	&  1130 	& 69 	&  65 
\\ 4592 	&  42.84 	&  -0.12 	 & 19:10:05.66   	&  08:45:0.0 	&  1.2 	&  880  	& 46 	&  40 
\\ 4532 	&  43.16 	&  0.85 	 & 19:07:10.12          &  09:29:0.0 	&  1.2 	&  2106 	& 248 	&  210 
\\ 4607 	&  43.41 	&  -0.03 	 & 19:10:49.45  	&  09:18:0.0 	&  1.5 	&  799  	& 94 	&  82 
\\ 4682 	&  43.66 	&  -0.91 	 & 19:14:28.73  	&  09:07:0.0 	&  1.5 	&  2782 	& 150 	&  123 
\\ 4623 	&  43.98 	&  0.07 	 & 19:11:33.25  	&  09:51:0.0 	&  1.3 	&  1125 	& 107 	&  69 
\\ 4590 	&  44.14 	&  0.55 	 & 19:10:05.41   	&  10:13:0.0 	&  1.1 	&  2153 	& 335 	&  314 
\\ 4699 	&  44.24 	&  -0.82 	 & 19:15:12.66  	&  09:40:0.0 	&  1.6 	&  1366 	& 99 	&  89 
\\ 4605 	&  44.71 	&  0.65 	 & 19:10:49.07  	&  10:46:0.0 	&  1.5 	&  3485 	& 436 	&  381 
\\ 4714 	&  44.81 	&  -0.72 	 & 19:15:56.63  	&  10:13:0.0 	&  1.5 	&  1469 	& 136 	&  118 
\\ 4683 	&  44.96 	&  -0.23 	 & 19:14:28.75  	&  10:35:0.0 	&  1.6 	&  1696 	& 151 	&  84 
\\ 4620 	&  45.28 	&  0.75 	 & 19:11:32.73  	&  11:19:0.0 	&  1.2 	&  4395 	& 392 	&  325 
\\ 4665 	&  45.69 	&  0.35 	 & 19:13:44.59  	&  11:30:0.0 	&  1.5 	&  1122 	& 167 	&  156 
\\ 4635 	&  45.85 	&  0.84 	 & 19:12:16.41  	&  11:52:0.0 	&  1.2 	&  4230 	& 584 	&  334 
\\ 4679 	&  46.26 	&  0.45 	 & 19:14:28.38  	&  12:03:0.0 	&  1.2 	&  1321 	& 265 	&  258 
\\ 4649 	&  46.42 	&  0.94 	 & 19:12:59.97  	&  12:25:0.0 	&  1.7 	&  2946 	& 380 	&  209 
\\ 4802 	&  46.93 	&  -0.82 	 & 19:20:20.63  	&  12:03:0.0 	&  1.7 	&  1464 	& 146 	&  99 
\\ 4663 	&  46.99 	&  1.04 	 & 19:13:43.53  	&  12:58:0.0 	&  1.6 	&  1943 	& 191 	&  76 
\\ 4817 	&  47.50 	&  -0.72 	 & 19:21:04.70  	&  12:36:0.0 	&  1.8 	&  1043 	& 93 	&  88 
\\ 4831 	&  48.07 	&  -0.62 	 & 19:21:48.80  	&  13:09:0.0 	&  1.8 	&  1792 	& 22 	&  18 
\\ 4801 	&  48.23 	&  -0.13 	 & 19:20:20.42  	&  13:31:0.0 	&  1.8 	&  1927 	& 212 	&  195 
\\ 4816 	&  48.80 	&  -0.03 	 & 19:21:04.48     	&  14:04:0.0 	&  1.8 	&  2088 	& 184 	&  155 
\\ 4859 	&  49.21 	&  -0.42 	 & 19:23:17.20  	&  14:15:0.0 	&  1.6 	&  1736 	& 165 	&  151 
\\ 4830 	&  49.36 	&  0.07 	 & 19:21:48.57  	&  14:37:0.0 	&  1.8 	&  1660 	& 5 	&  0 
\\ 4902 	&  49.63 	&  -0.80 	 & 19:25:30.00  	&  14:26:0.0 	&  1.8 	&  2419 	& 199 	&  184 
\\ 4916 	&  50.19 	&  -0.70 	 & 19:26:14.31  	&  14:59:0.0 	&  0.9 	&  2174 	& 233 	&  218 
\\ 4857 	&  50.50 	&  0.28 	 & 19:23:16.58  	&  15:43:0.0 	&  1.6 	&  1983 	& 143 	&  133 
\\ 4930 	&  50.76 	&  -0.59 	 & 19:26:58.69  	&  15:32:0.0 	&  0.9 	&  2331 	& 92 	&  80 
\\ 4901 	&  50.92 	&  -0.10 	 & 19:25:29.72  	&  15:54:0.0 	&  1.6 	&  1983 	& 120 	&  112 
\\ 4973 	&  51.18 	&  -0.97 	 & 19:29:11.95  	&  15:43:0.0 	&  0.9 	&  1460 	& 142 	&  93 
\\ 4944 	&  51.33 	&  -0.49 	 & 19:27:43.14  	&  16:05:0.0 	&  1.1 	&  1946 	& 118 	&  109 
\\ 4915 	&  51.48 	&  0.00 	 & 19:26:14.02  	&  16:27:0.0 	&  0.9 	&  2384 	& 236 	&  228 
\\ 4986 	&  51.75 	&  -0.87 	 & 19:29:56.62  	&  16:16:0.0 	&  0.9 	&  1695 	& 59 	&  55 
\\ 4928 	&  52.05 	&  0.11 	 & 19:26:58.16  	&  17:00:0.0 	&  0.9 	&  1984 	& 255 	&  139 
\\ 4899 	&  52.20 	&  0.60 	 & 19:25:28.54  	&  17:22:0.0 	&  1.7 	&  1995 	& 156 	&  139 
\\ 5000 	&  52.32 	&  -0.76 	 & 19:30:41.19  	&  16:49:0.0 	&  1.3 	&  2357 	& 207 	&  196 
\\ 4972 	&  52.47 	&  -0.27 	 & 19:29:11.83  	&  17:11:0.0 	&  1.2 	&  1283 	& 86 	&  82 
\\ 4942 	&  52.62 	&  0.22 	 & 19:27:42.38  	&  17:33:0.0 	&  0.9 	&  1687 	& 118 	&  112 
\\ 4912 	&  52.77 	&  0.70 	 & 19:26:12.55  	&  17:55:0.0 	&  0.9 	&  2044 	& 190 	&  162 
\\ 5015 	&  52.88 	&  -0.65 	 & 19:31:25.81  	&  17:22:0.0 	&  1.2 	&  2380 	& 174 	&  124 
\\ 4985 	&  53.03 	&  -0.16 	 & 19:29:56.29  	&  17:44:0.0 	&  0.9 	&  1008 	& 35 	&  35 
\\ 4926 	&  53.34 	&  0.81 	 & 19:26:56.63  	&  18:27:59.9 	&  0.9 	&  2075 	& 147 	&  130 
\\ 4999 	&  53.60 	&  -0.05 	 & 19:30:40.82  	&  18:17:0.0 	&  1.3 	&  1364 	& 118 	&  111 
\\ 4970 	&  53.75 	&  0.43 	 & 19:29:10.98  	&  18:38:59.9 	&  1.6 	&  1694 	& 171 	&  138 
\\ 4939 	&  53.90 	&  0.92 	 & 19:27:40.50  	&  19:00:59.9 	&  1.1 	&  3148 	& 402 	&  299 
\\ 5014 	&  54.17 	&  0.06 	 & 19:31:25.45  	&  18:49:59.9 	&  1.3 	&  3285 	& 324 	&  219 
\\ 4983 	&  54.32 	&  0.54 	 & 19:29:55.14  	&  19:11:59.9 	&  0.9 	&  5067 	& 188 	&  88 
\\ 4997 	&  54.89 	&  0.66 	 & 19:30:39.36  	&  19:44:59.9 	&  1.3 	&  5019 	& 436 	&  192 
\\ 5098 	&  55.00 	&  -0.70 	 & 19:35:54.90  	&  19:11:59.9 	&  0.9 	&  3108 	& 287 	&  108 
\\ 5069 	&  55.15 	&  -0.21 	 & 19:34:24.73  	&  19:33:59.9 	&  0.8 	&  4650 	& 400 	&  224 
\\ 5041 	&  55.30 	&  0.28 	 & 19:32:54.41  	&  19:55:59.9 	&  0.9 	&  4928 	& 183 	&  98 
\\ 5112 	&  55.57 	&  -0.58 	 & 19:36:39.94  	&  19:44:59.9 	&  1.1 	&  4229 	& 619 	&  397 
\\ 5083 	&  55.72 	&  -0.09 	 & 19:35:09.55   	&  20:06:59.9 	&  0.9 	&  4422 	& 561 	&  379 
\\ 5025 	&  56.02 	&  0.88 	 & 19:32:08.01   	&  20:50:59.9 	&  1.1 	&  2828 	& 60 	&  30 
\\ 5126 	&  56.14 	&  -0.47 	 & 19:37:25.09  	&  20:17:59.9 	&  0.9 	&  4457 	& 399 	&  240 
\\ 5097 	&  56.28 	&  0.02 	 & 19:35:54.46  	&  20:39:59.9 	&  0.9 	&  2286 	& 139 	&  131 
\\ 5039 	&  56.58 	&  1.00 	 & 19:32:52.14  	&  21:23:59.9 	&  0.9 	&  2723 	& 190 	&  112 
\\ 5140 	&  56.70 	&  -0.35 	 & 19:38:10.34  	&  20:50:59.9 	&  1.6 	&  2467 	& 198 	&  186 
\\ 5111 	&  56.85 	&  0.14 	 & 19:36:39.48  	&  21:12:59.9 	&  1.0 	&  1207 	& 70 	&  67 
\\ 5080 	&  57.00 	&  0.62 	 & 19:35:08.16   	&  21:34:59.9 	&  0.8 	&  1734 	& 168 	&  53 
\\ 5124 	&  57.41 	&  0.25 	 & 19:37:24.30  	&  21:45:59.9 	&  1.0 	&  689  	& 79 	&  78 
\\ 5095 	&  57.56 	&  0.74 	 & 19:35:52.71  	&  22:07:59.9 	&  0.9 	&  1645 	& 102 	&  97 
\\ 5138 	&  57.98 	&  0.37 	 & 19:38:09.21   	&  22:18:59.9 	&  1.5 	&  1558 	& 85 	&  47 
\\ 5108 	&  58.13 	&  0.86 	 & 19:36:37.36  	&  22:40:59.9 	&  0.9 	&  2288 	& 178 	&  106 
\\ 5152 	&  58.54 	&  0.49 	 & 19:38:54.23  	&  22:51:59.9 	&  1.5 	&  1867 	& 133 	&  106 
\\ 5262 	&  59.24 	&  -0.73 	 & 19:45:00.13   	&  22:51:59.9 	&  1.1 	&  3095 	& 240 	&  221 
\\ 5277 	&  59.80 	&  -0.61 	 & 19:45:45.96  	&  23:24:59.9 	&  1.1 	&  2592 	& 207 	&  195 
\\ 5261 	&  60.51 	&  0.01 	 & 19:44:59.57  	&  24:19:59.9 	&  1.3 	&  4143 	& 413 	&  341 
\\ 5275 	&  61.07 	&  0.13 	 & 19:45:45.38  	&  24:52:59.9 	&  1.1 	&  3989 	& 387 	&  261 
\\ 5273 	&  62.33 	&  0.88 	 & 19:45:42.67  	&  26:20:59.8 	&  1.1 	&  2978 	& 246 	&  194 
\\ 5285 	&  62.89 	&  1.01 	 & 19:46:27.69  	&  26:53:59.8 	&  1.0 	&  4273 	& 288 	&  116 
\\ 5489 	&  65.02 	&  -0.76 	 & 19:58:11.41  	&  27:48:59.8 	&  1.5 	&  2900 	& 196 	&  74 \\
\hline
\hline\\
 TOTAL 	&  	&  	 &  	&   	&  	&      	260223         &   23050  &  15312
\enddata
\end{deluxetable}

\clearpage

\begin{deluxetable}{ccccc}
\tabletypesize{\scriptsize}
\tablecaption{Survey observational properties\label{iphaszeropoints}}
\tablewidth{0pt}
\tablehead{ \colhead{Filter}    &    \colhead{Central wavelength } &   \colhead{Zeropoint}   &    \colhead{FWHM}   &   \colhead{ Lower flux limits}  \\
            \colhead{}     &    \colhead{[$\mu$m]}  &   \colhead{[Jy]}   &   \colhead{[arcseconds]}  &  \colhead{[mJy/mag]}  
     }   
\startdata
IPHAS r$^{\prime}$& 0.624   &3173.3      & $<2.0$      & 0.2/18          \\ 
IPHAS i$^{\prime}$& 0.774   &2515.7      & $<2.0$      & 0.02/20.5        \\
IPHAS H$\alpha$         & 0.656   &2974.4      & $<2.0$      & 0.02/20.5        \\
2MASS J           & 1.235   & 1594.5     & $\sim$3.1            & 0.7/15.8        \\
2MASS H           & 1.662   & 1024.5     & $\sim$3.0            & 0.9/15.1        \\
2MASS K           & 2.159   & 666.7    & $\sim$3.1            & 1.2/14.3        \\
GLIMPSE IRAC 1         & 3.550  &277.5 &1.6&0.6/14.2\\
GLIMPSE IRAC 2         & 4.493  &179.5 &1.7&0.4/14  \\
GLIMPSE IRAC 3         & 5.731  &116.5 &1.8&2/11.9  \\
GLIMPSE IRAC 4         & 7.872  &63.1 &1.9&10/9.5  \\
MIPSGAL MIPS 1         &23.670  &7.13 &5.5 & 1/9.6  \\
\hline
\enddata
\tablecomments{Filter characteristics, spatial resolution and photometric
sensitivities of the IPHAS, 2MASS, GLIMPSE and MIPSGAL 24-$\mu$m
surveys. Note that the GLIMPSE catalog includes sources with fluxes 
smaller than
these sensitivities, provided they have a SN$>3$ and that they meet the $2+1$ criterion in the
other bands. The zeropoint (luminosity of a 0th magnitude star) in each
band is given for conversion from fluxes to magnitudes. From
\cite{drew05}, \cite{cutri03}, the {\em Spitzer} Observer Manual
Version 7.1, the GLIMPSE Legacy Data Products Notes version
2.0 and the Infrared Processing and Analysis Center
(IPAC).}
\end{deluxetable}

\begin{deluxetable}{cccccc}
\tabletypesize{\scriptsize}
\tablecaption{Source counts and mean observed  magnitude for the IPHAS-GLIMPSE sample of A-type
stars\label{gliphas}}
\tablewidth{0pt}
\tablehead{ \colhead{Band}    &    \colhead{Number of stars}  &   \colhead{Mean observed  magnitude}         }
\startdata
r$^{\prime}$    & 15312& 16.179   \\
i$^{\prime}$    & 15312& 14.892   \\
J    & 12748& 14.045  \\
H    & 12485& 13.546 \\
K    & 11198& 13.281 \\
3.6  & 11175& 12.982   \\
4.5  & 11037& 12.969 \\
5.8  &  5111& 12.385 \\
8    &  2751& 12.097 \\
\hline
\enddata

\end{deluxetable}

\begin{figure}
\begin{center}
\includegraphics[scale=0.5,angle=0,clip=true]
{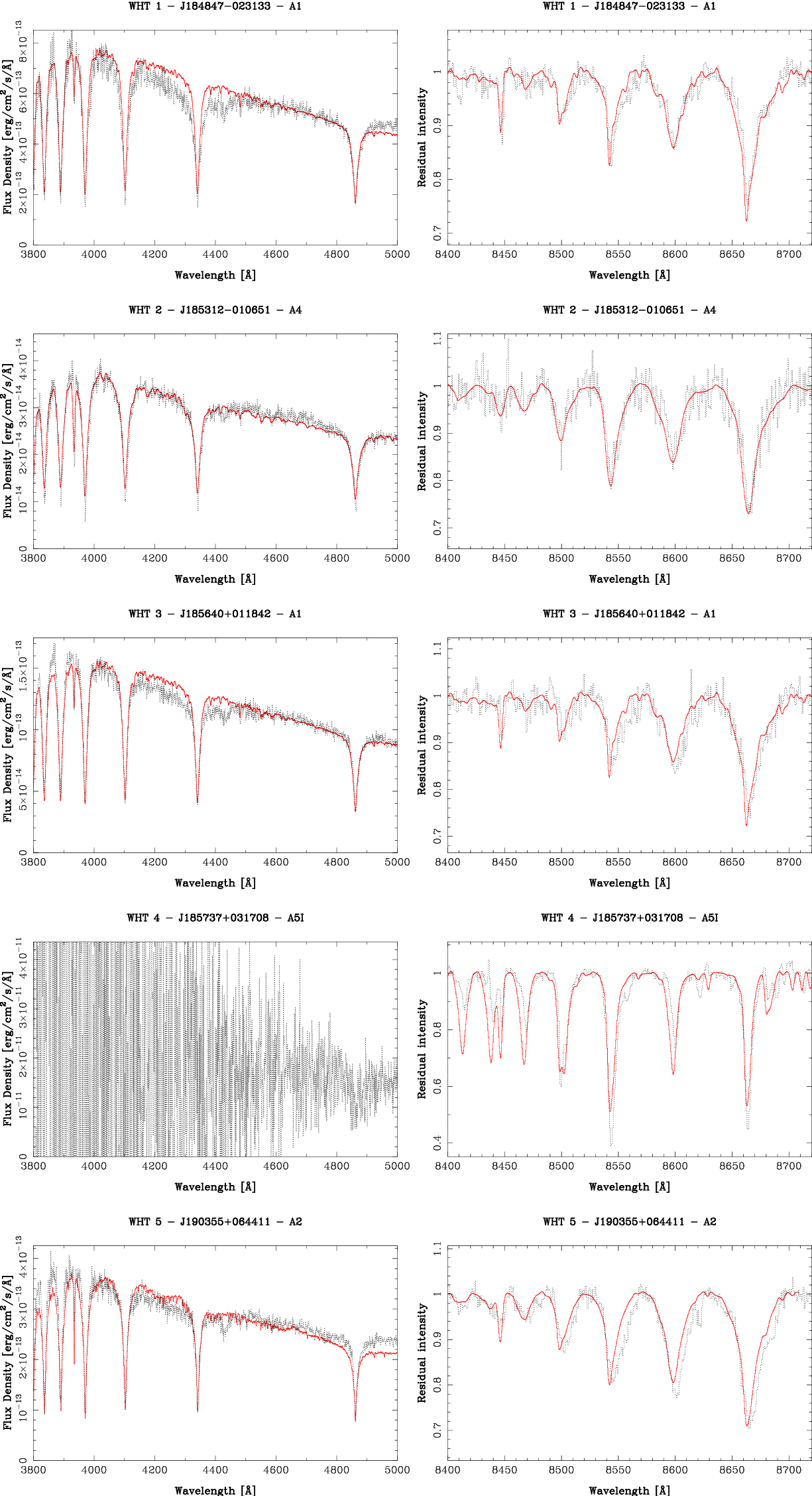}
\caption{WHT ISIS Spectra of IPHAS-selected A-type stars. The
dereddened flux-calibrated spectra of the targets (black dotted lines)
are shown for the blue spectral region (left-hand panels) and for the
red spectral region (right-hand panels), and are compared to the
spectra of the A-type spectral standards that provide the best match
to their red spectral features (red solid lines).  The spectral
subtypes of these spectral standards are shown at the top of each
plot. All of the plotted spectral standards are luminosity class V
dwarfs, except for WHT~4, which is best matched by an A5Ia supergiant
spectrum.\label{wht}}
\end{center}
\end{figure}

\addtocounter{figure}{-1}
\begin{figure}
\begin{center}
\includegraphics[scale=0.5,angle=0,clip=true]
{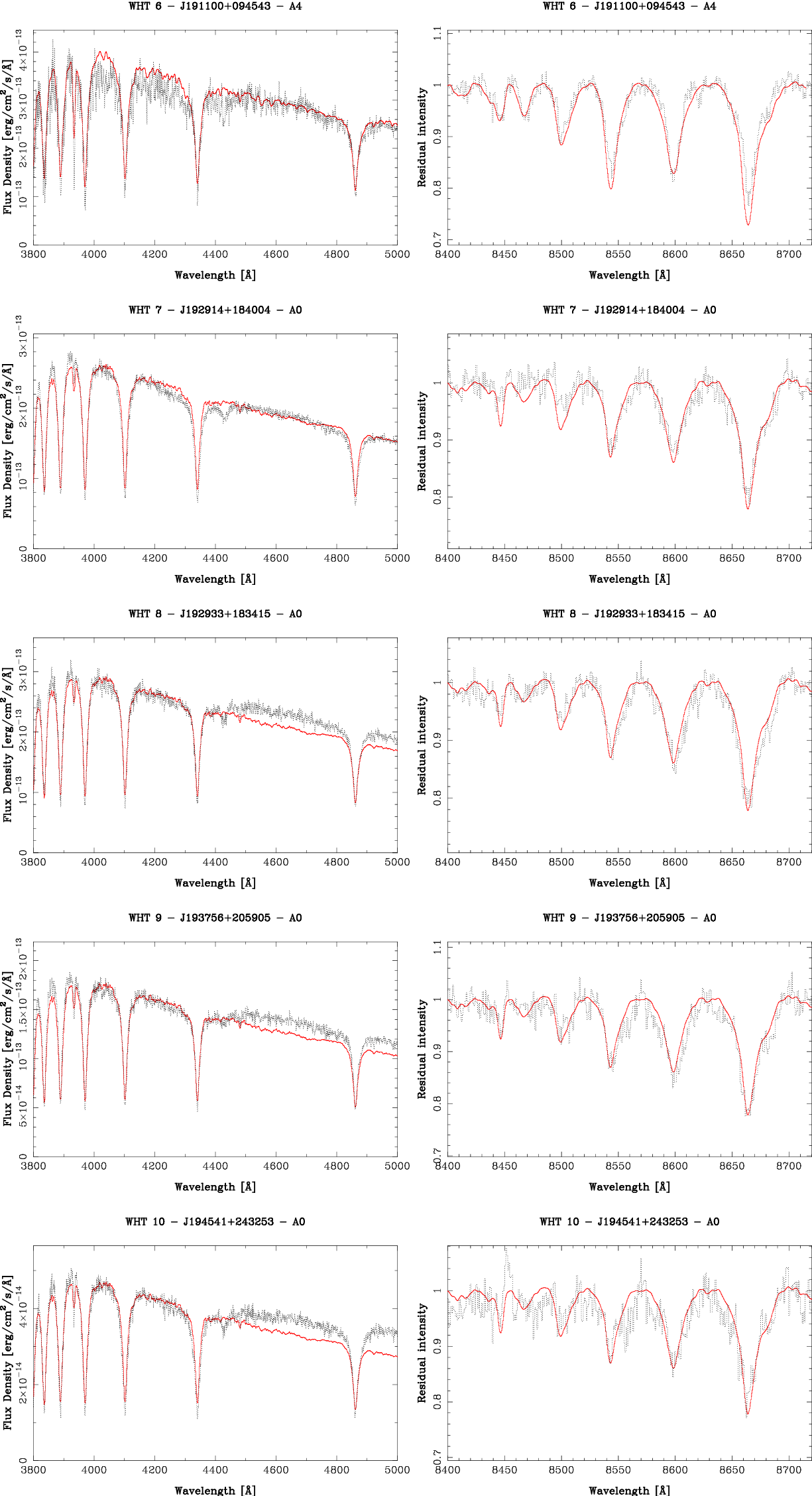}
\caption{{\it{continued}}}\label{wht}
\end{center}
\end{figure}

\begin{figure}
\begin{center}
\includegraphics[scale=0.6,angle=0,clip=true]
{f4.eps}
\caption{Dereddened $(J-H)-(K-8)$ color-color diagram for the $2692$
A-type stars in the GLIPHAS sample.  The 20 objects having a $(K-8)$
color excess with SN$>3$ are highlighted with diamond symbols.
}\label{fig:k8}
\end{center}
\end{figure}

\begin{figure}
\begin{center}
\includegraphics[scale=0.8,angle=0,clip=true]
{f5.eps}
\caption{
{\it{Left-hand panel}}: Distributions of $E(K-8)$ color
 excesses for $2692$ A-type stars in the GLIPHAS
 sample. {\it{Right-hand panel}}: Distributions of signal-to-noise for
$(K-8)$ color excesses.
 }\label{fig:distributions}
\end{center}
\end{figure}

\begin{figure}
\begin{center}
\includegraphics[scale=0.5,angle=0,clip=true]
{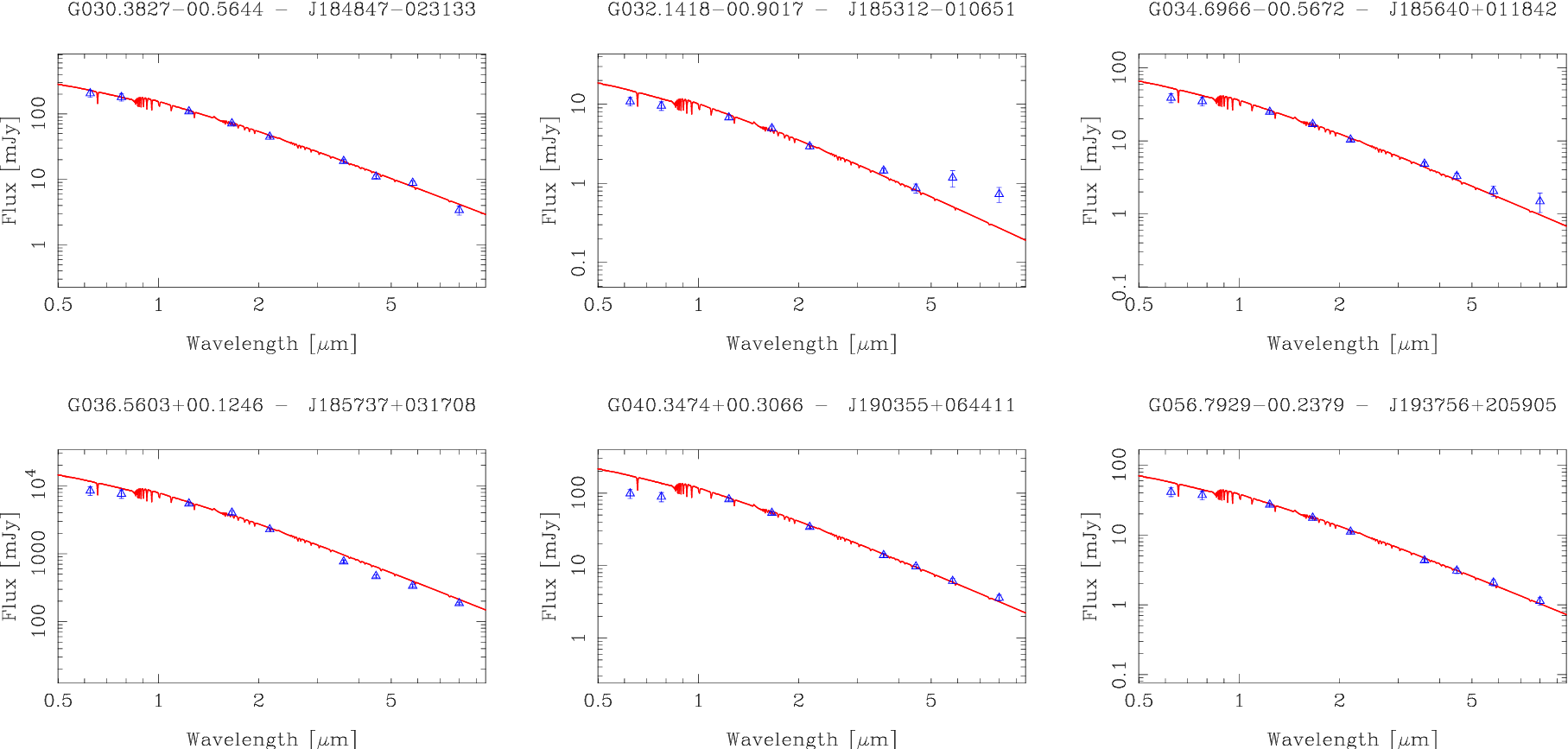}
\caption{ Dereddened optical IPHAS, 2MASS and GLIMPSE photometry for
the 6 WHT A-type stars with no evidence of 8-$\mu$m excesses (blue
triangles with error bars denoting the 1$\sigma$ photometric error of
each measurement). For each star, the red solid line corresponds to
the photospheric reference of an A3V SED, normalised to the K-band
flux.
\label{fig:sedwht}}
\end{center}
\end{figure}

\begin{deluxetable}{ccccccc}
\tabletypesize{\scriptsize}
\tablecaption{WHT/ISIS Spectral Classifications\label{speclass}}
\tablewidth{0pt}
\tablehead{ \colhead{}    &    \colhead{}  &  \colhead{}  & \colhead{}  & \colhead{Ca~{\sc ii} EW's (\AA )}   & \colhead{} & \colhead{} \\ 
            \colhead{Star}&    \colhead{Red-based}   &  \colhead{Blue-Based}     & \colhead{8498.02~\AA }  & \colhead{8542.09~\AA }   & 
\colhead{8662.14~\AA } & \colhead{3933.663~\AA }} 
\startdata

WHT1 - J184847-023133   & A1V   &      A1V & 0.64(0.21)   & 1.71(0.27)   &   3.65(0.32)  & 1.36(0.31) \\	
WHT2 - J185312-010651   & A4V   &      A6V & 0.91(0.32)   & 1.83(0.39)   &    4.39(0.31)  & 1.44(0.42) \\
WHT3 - J185640+011842   & A1V   &      A2V & 0.52(0.25)   & 3.02(0.29)   &   4.87(0.39)  & 0.87(0.21) \\	
WHT4 - J185737+031708   & A5Ia  &      -   & 2.76(0.32)   & 5.48(0.41)   &   4.41(0.37)  & - \\
WHT5 - J190355+064411   & A2V   &      A3V & 1.33(0.19)   & 2.95(0.22)   &   6.17(0.29)  & 1.08(0.21) \\	
WHT6 - J191100+094543   & A4V   &      A7V & 1.15(0.34)   & 2.36(0.31)   &   4.09(0.32)  & 3.34(0.37) \\			
WHT7 - J192914+184004   & A0V   &      A0V & 0.36(0.17)   & 1.44(0.21)   &   4.41(0.27)  & 0.61(0.19) \\			
WHT8 - J192933+183415   & A0V   &      A0V & 0.33(0.17)   & 1.72(0.23)   &   4.55(0.28)  & 0.44(0.24) \\	
WHT9 - J193756+205905   & A0V   &      A2V & 0.49(0.19)   & 1.68(0.24)   &   4.31(0.33)  & 1.05(0.25) \\		
WHT10 - J194541+243253  & A0V   &      A2V & 0.25(0.19)   & 1.51(0.34)   &   3.36(0.41)  & 1.08(0.33) \\

\hline
\enddata
\tablecomments{Blue- (Ca~{\sc ii}~K) and red-based (Ca~{\sc ii} 
IR-triplet) spectral
classifications for stars observed with the WHT.  The red-based spectral 
types   
are 0.9 sub-types earlier on average than the blue-based ones (see text). 
The measured equivalent widths of the Ca~{\sc ii}-K and
IR-triplet lines are given in $\rm{\AA}$, with the numbers
in brackets denoting the 2$\sigma$ uncertainties.}
\end{deluxetable}


\begin{deluxetable}{cccccccccccccc}
\tabletypesize{\scriptsize}
\rotate
\tablecaption{Photometry for GLIPHAS A-type stars with WHT spectra but with no 8-$\mu$m
excesses\label{whtphotom}}
\tablewidth{0pt}
 \setlength{\tabcolsep}{0.4mm}
\tablehead{
\colhead{WHT} &\colhead{ IPHAS}      &\colhead{   
r$^{\prime}$}   &\colhead{   r$^{\prime}$-i$^{\prime}$}     &\colhead{J}   &\colhead{  J-H}   &\colhead{  J-K} &\colhead{ J-[3.6]}&\colhead{ [3.6]}&\colhead{ [3.6]-[4.5]} &\colhead{ [3.6]-[5.8]} &\colhead{  [3.6]-[8.0]}  & \colhead{    A$_{r}$} 
&\colhead{ d }  \\
\colhead{No.}    &\colhead{          ID} &\colhead{      (mag)}  &\colhead{      (mag)}              &\colhead{        (mag)}      &\colhead{ (mag)}  &\colhead{ (mag)}&\colhead{  (mag)} &\colhead{ (mag)}&\colhead{   (mag)}     &\colhead{   (mag)}     &\colhead{     (mag)}  &\colhead{     (mag)}   & \colhead{(kpc)}   
}
\startdata

 1& J184847-023133  & 14.48(0.02)  & 1.09(0.03)& 11.69(0.02)& 0.44(0.03) &0.70(0.03) &0.97(0.05) &10.72& -0.05(0.08) &0.18(0.09) &0.79(0.08) &  4.1 (0.3) & 0.6 (0.1) \\  
 2& J185312-010651  & 15.70(0.01)  & 0.57(0.01)& 14.00(0.05)& 0.32(0.08) &0.35(0.09) &0.65(0.09) &13.35& -0.05(0.16) &0.75(0.26) &0.90(0.24) &  2.0 (0.2) & 2.6 (0.5)   \\ 
 3& J185640+011842  & 14.71(0.02)  & 0.76(0.03)& 12.85(0.02)& 0.33(0.03) &0.47(0.04) &0.75(0.06) &12.10& 0.10(0.12) &0.45(0.13) &0.93(0.20)  &  2.8 (0.3) & 1.2 (0.2)  \\ 
 4& J185737+031708  & 15.19(0.01)  & 2.38(0.01)& 9.18(0.02)& 1.13(0.05) &1.70(0.03) &2.05(0.04) &7.13& 0.10(0.05) &0.21(0.05) &0.24(0.04)    &  9.1 (0.2) & 0.1 (0.0)  \\ 
 5& J190355+064411  & 14.24(0.04)  & 0.90(0.06)& 11.74(0.02)& 0.34(0.04) &0.57(0.03) &0.74(0.07) &11.00& 0.14(0.09) &0.11(0.10) &0.53(0.09)  &  3.3 (0.3) & 0.7 (0.2)   \\
 9& J193756+205905  & 14.46(0.02)  & 0.71(0.03)& 12.70(0.02)& 0.27(0.03) &0.42(0.03) &0.49(0.05) &12.21& 0.15(0.10) &0.19(0.12) &0.78(0.09)  &  2.6 (0.3) & 1.1 (0.2)

\enddata \tablecomments{Observed magnitudes and colors in the
different IPHAS, 2MASS and IRAC bands, together with visual
extinctions, A$_{\rm r^{\prime}}$, and spectrophotometric distances
derived from the IPHAS colors. The figures in brackets denote the
1~$\sigma$ uncertainties. For each star, extinctions and distances
were computed assuming all possible spectral types (A0-5), and the rms
value of each quantity was added quadratically to the photometric errors to derive
the final uncertainties quoted here.
}
\end{deluxetable}
\clearpage

\begin{deluxetable}{cccccccccccccc}
\tabletypesize{\scriptsize}
\rotate
\tablecaption{Photometry for IPHAS A-type stars with 8-$\mu$m
excesses\label{24excess}}
\tablewidth{0pt}
 \setlength{\tabcolsep}{0.4mm}
\tablehead{
 \colhead{GLIPHAS} &\colhead{ IPHAS}      &\colhead{   
r$^{\prime}$}   &\colhead{   r$^{\prime}$-i$^{\prime}$}  &\colhead{  r$^{\prime}$-J}   &\colhead{J}   &\colhead{  J-H}   &\colhead{  J-K} &\colhead{ J-[3.6]}&\colhead{ [3.6]}&\colhead{ [3.6]-[4.5]} &\colhead{ [3.6]-[5.8]} &\colhead{  [3.6]-[8.0]}  &\colhead{[24]} \\
     \colhead{No.}    &\colhead{          ID} &\colhead{      (mag)}  &\colhead{      (mag)}     &\colhead{            (mag)}             &\colhead{        (mag)}      &\colhead{ (mag)}  &\colhead{ (mag)}&\colhead{  (mag)} &\colhead{ (mag)}&\colhead{   (mag)}     &\colhead{   (mag)}     &\colhead{     (mag)}   &\colhead{ (mag)} 
}
\startdata

   1 & J190602+073418  & 15.20(0.02)  & 1.07(0.03)& 2.16(0.03) &13.04(0.02)& 0.40(0.04) &0.57(0.04) &0.73(0.06) &12.31& 0.10(0.12) &0.10(0.14) &1.30(0.11)   &  7.93 0.11 \\ 
   2 & J190650+090108  & 16.89(0.01)  & 1.60(0.01)& 3.44(0.03) &13.45(0.03)& 0.57(0.04) &0.91(0.05) &1.18(0.11) &12.27& 0.46(0.17) &0.27(0.16) &2.84(0.18)   & $<$ \\ 
   3 & J190952+070514  & 17.41(0.02)  & 1.45(0.02)& 3.80(0.03) &13.61(0.02)& 0.62(0.03) &0.93(0.03) &1.33(0.06) &12.28& -0.37(0.14) &0.06(0.17) &1.94(0.29)   & $<$ \\ 
   4 & J191100+094543  & 15.18(0.01)  & 1.18(0.02)& 3.00(0.02) &12.18(0.02)& 0.57(0.03) &1.04(0.03) &1.98(0.04) &10.20& 0.55(0.07) &1.05(0.06) &1.82(0.05)   & 7.06 0.11 \\ 
   5 & J191512+120907  & 17.54(0.04)  & 1.19(0.05)& 2.89(0.06) &14.65(0.05)& 0.58(0.07) &0.82(0.08) &0.89(0.08) &13.76& 0.07(0.17) &1.13(0.25) &2.06(0.14)   & $<$ \\ 
   6 & J191947+132707  & 17.82(0.00)  & 1.30(0.01)& 3.33(0.03) &14.49(0.03)& 0.62(0.07) &0.95(0.09) &0.97(0.08) &13.52& 0.23(0.18) &2.28(0.18) &2.76(0.15)   & $<$ \\ 
   7 & J192227+141309  & 14.83(0.04)  & 0.95(0.05)& 2.35(0.04) &12.48(0.02)& 0.34(0.04) &0.55(0.03) &0.82(0.06) &11.66& 0.02(0.09) &0.65(0.18) &1.52(0.33)   & $<$ \\ 
   8 & J192612+163544  & 16.87(0.03)  & 0.94(0.04)& 2.44(0.04) &14.43(0.03)& 0.37(0.05) &0.72(0.06) &0.82(0.10) &13.61& 0.08(0.17) &0.62(0.32) &1.31(0.24)   & $<$ \\ 
   9 & J192914+184004  & 14.53(0.02)  & 0.85(0.03)& 2.12(0.03) &12.41(0.02)& 0.36(0.03) &0.52(0.03) &0.68(0.05) &11.73& 0.33(0.09) &0.66(0.10) &2.06(0.07)   & 6.40 0.11 \\ 
   10 & J192933+183415  & 14.15(0.01)  & 0.68(0.01)& 1.81(0.02) &12.34(0.02)& 0.42(0.03) &0.77(0.03) &1.28(0.06) &11.06& 0.19(0.07) &0.38(0.08) &1.29(0.06)   & 7.61 0.10 \\ 
   11 & J193110+164225  & 17.54(0.00)  & 1.35(0.01)& 3.22(0.03) &14.32(0.03)& 0.45(0.05) &0.79(0.04) &0.95(0.07) &13.37& 0.14(0.14) &0.98(0.27) &1.67(0.19)   & $<$ \\ 
   12 & J193117+185048  & 17.34(0.03)  & 1.13(0.04)& 2.79(0.04) &14.55(0.03)& 0.46(0.05) &0.57(0.06) &1.04(0.08) &13.51& 0.24(0.15) &1.22(0.16) &1.59(0.17)   & $<$ \\ 
   13 & J193126+184131  & 15.74(0.00)  & 0.99(0.01)& 2.48(0.02) &13.26(0.02)& 0.47(0.03) &0.79(0.03) &0.82(0.06) &12.44& 0.05(0.11) &0.56(0.14) &1.53(0.18)   & $<$ \\ 
   14 & J193259+195709  & 16.77(0.02)  & 1.08(0.03)& 2.65(0.03) &14.12(0.02)& 0.59(0.03) &0.95(0.03) &1.55(0.09) &12.57& 0.22(0.15) &1.12(0.14) &2.78(0.12)   & $<$ \\ 
   15 & J193516+195803  & 15.97(0.00)  & 0.77(0.01)& 1.94(0.02) &14.03(0.02)& 0.34(0.04) &0.49(0.05) &0.58(0.07) &13.45& 0.04(0.13) &0.51(0.32) &1.58(0.22)   & $<$ \\ 
   16 & J193555+194654  & 17.42(0.01)  & 1.03(0.02)& 2.78(0.03) &14.64(0.03)& 0.42(0.05) &0.56(0.07) &0.87(0.08) &13.77& 0.03(0.17) &1.10(0.29) &1.33(0.28)   & $<$ \\ 
   17 & J194541+243253  & 15.20(0.10)  & 0.54(0.14)& 1.53(0.11) &13.67(0.03)& 0.23(0.05) &0.38(0.05) &0.69(0.07) &12.98& 0.44(0.12) &1.11(0.14) &1.57(0.09)   & 8.36 0.15 \\ 
 

\enddata
\tablecomments{Observed magnitudes and colors in the different IPHAS, 
2MASS and IRAC bands, together with our measured 24-$\mu$m fluxes from 
MIPSGAL images. The figures in brackets denote the 1~$\sigma$ uncertainties.}
\end{deluxetable}
\clearpage

\begin{deluxetable}{ccccccccccccc}
\tabletypesize{\scriptsize}
\rotate
\tablecaption{Derived parameters for IPHAS A-type stars with 8-$\mu$m
excesses.\label{derived24}}
\tablewidth{0pt}
 \setlength{\tabcolsep}{0.9mm}
\tablehead{
 \colhead{GLIPHAS}&\colhead{IPHAS}&\colhead{ GLIMPSE} &\colhead{    A$_{r}$} 
&\colhead{ d }   &\colhead{  3.6$\mu$m}  &\colhead{  4.5$\mu$m} &\colhead{ 5.8$\mu$m}   &\colhead{    8$\mu$m}  &\colhead{ 24$\mu$m}   &\colhead{  T}  &\colhead{ L$_{\rm IR}$/L$_{\star}$}  \\
 \colhead{    No.}   &\colhead{ ID}&\colhead{  ID}    &\colhead{   (mag)}    &\colhead{(kpc)}   &\colhead{  (mJy)}&\colhead{ (mJy)}&\colhead{ (mJy)} &\colhead{ (mJy)} &\colhead{ (mJy)}&\colhead{ (K)} &\colhead{}  
}                     
\startdata

1 & J190602+073418  & G041.3305+00.2230 &  4.0 (0.2) & (0.8 0.2) &  -         &  -        & -          & 2.0 (7.0 )& 4.8 (9.0)  & 283 (24)    & 2.2e-03 \\ 
2 & J190650+090108  & G042.7075+00.7117 &  6.0 (0.2) & (0.7 0.1) &  -         &  -        & -          &13.7 (6.0 )& -          & $\leq$211   & -       \\ 
3 & J190952+070514  & G041.3387-00.8442 &  5.5 (0.2) & (1.2 0.2) &  -         &  -        & -          & 5.3 (3.1 )& -          & $\leq$194   & -       \\ 
4 & J191100+094543  & G043.8401+00.1431 &  4.4 (0.2) & (0.7 0.1) &14.3 (10.3) & 20.1(11.8)& 25.7 (18.5)&32.4 (33.5)& 10.4(8.9)  & 654 (33)    & 8.5e-03 \\ 
5 & J191512+120907  & G046.4350+00.3389 &  4.5 (0.3) & (2.0 0.4) &  -         &  -        &-           & 1.4 (7.1 )& -          &$\leq$243 (  & -       \\   
6 & J191947+132707  & G048.1073-00.0432 &  4.9 (0.2) & (1.9 0.4) &  -         &  -        & 4.0   (5.3)& 3.7 (7.1 )& -          &  271 (63)   & 1.9e-02 \\ 
7 & J192227+141309  & G049.0889-00.2538 &  3.5 (0.3) & (0.9 0.2) &  -         &  -        & -          & 5.1 (3.0 )& $<$87.3    & $\leq$ 299  & -       \\ 
8 & J192612+163544  & G051.6091+00.0765 &  3.5 (0.3) & (2.3 0.5) & -          &  -        & -          & 0.6 (3.2 )& -          &$\leq$257    & -       \\ 
9 & J192914+184004  & G053.7744+00.4315 &  3.1 (0.2) & (0.9 0.2) & -          &           & 2.1 (4.4)  & 8.6 (14.9)& 19.9(9.2)  &298 (20)     & 6.6e-03 \\
10 & J192933+183415  & G053.7251+00.3191 &  2.5 (0.2) & (1.0 0.2) & 3.3 (5.0) & 3.1 (7.4) & 3.4 (7.8)  & 7.0 (20.7)& 15.6(9.6)  & 497 (57)    & 3.0e-03 \\ 
11 & J193110+164225  & G052.2752-00.9138 &  5.1 (0.2) & (1.5 0.3) & -         &  -        & -          & 1.3 (4.3 )& $<$5.4     & $\leq$245   & -      \\ 
12 & J193117+185048  & G054.1653+00.0894 &  4.2 (0.3) & (2.0 0.4) & -         &    -      & 1.3 (5.2)  & 1.1 (5.7 )& $<$22.8    &455  (20)    & 4.3e-03 \\   
13 & J193126+184131  & G054.0455-00.0137 &  3.7 (0.2) & (1.2 0.3) & -         &  -        & -          & 2.3 (4.4 )& -          &   $\leq$183 & -       \\   
14 & J193259+195709  & G055.3273+00.2720 &  4.0 (0.3) & (1.7 0.4) & 0.9 (3.2) &  0.9 (3.0) & 2.8 (6.8) & 8.9 (11.3)& $<$46.9    &342  (43)    & 1.6e-02 \\   
15 & J193516+195803  & G055.6008-00.1905 &  2.9 (0.2) & (2.0 0.4) & -         &  -        & -          & 1.0 (3.7 )& -          &   $\leq$168 & -       \\ 
16 & J193555+194654  & G055.5133-00.4151 &  3.8 (0.2) & (2.5 0.5) & -         &  -        &  -         & 0.7 (3.0 )& $<$0.8     &  $\leq$ 367 & -       \\ 
17 & J194541+243253  & G060.7727-00.0227 &  1.9 (0.6) & (2.2 0.7) &   -        & 0.8 (4.1) & 1.6 (6.1) & 1.5 (14.0)& 3.3 (6.3)  &    482 (15) & 3.8e-03 \\

 \enddata \tablecomments{ Derived parameters for stars with mid-IR
excesses with respect to the SED of an A3V star normalised to the
dereddened K-band fluxes. The excess fluxes listed for each wavelength
are in mJy. The numbers in brackets denote the signal-to-noise ratios
of the excess flux detections, except for the last column, where they
represent 1~$\sigma$ uncertainties. Only detections with S/N ratios
larger than 3.0 are shown.  For each star, extinctions and distances
were computed assuming all possible spectral types (A0-5), and the RMS
value of each quantity was added quadratically to the photometric
errors to derive the final uncertainties quoted here. For sources that
show excess fluxes at more than one wavelength the infrared excess
L$_{\rm IR}$/L$_{\star}$ are shown.}
\end{deluxetable}
\clearpage

\begin{figure}
\begin{center}
\includegraphics[scale=0.5,angle=0,clip=true]
{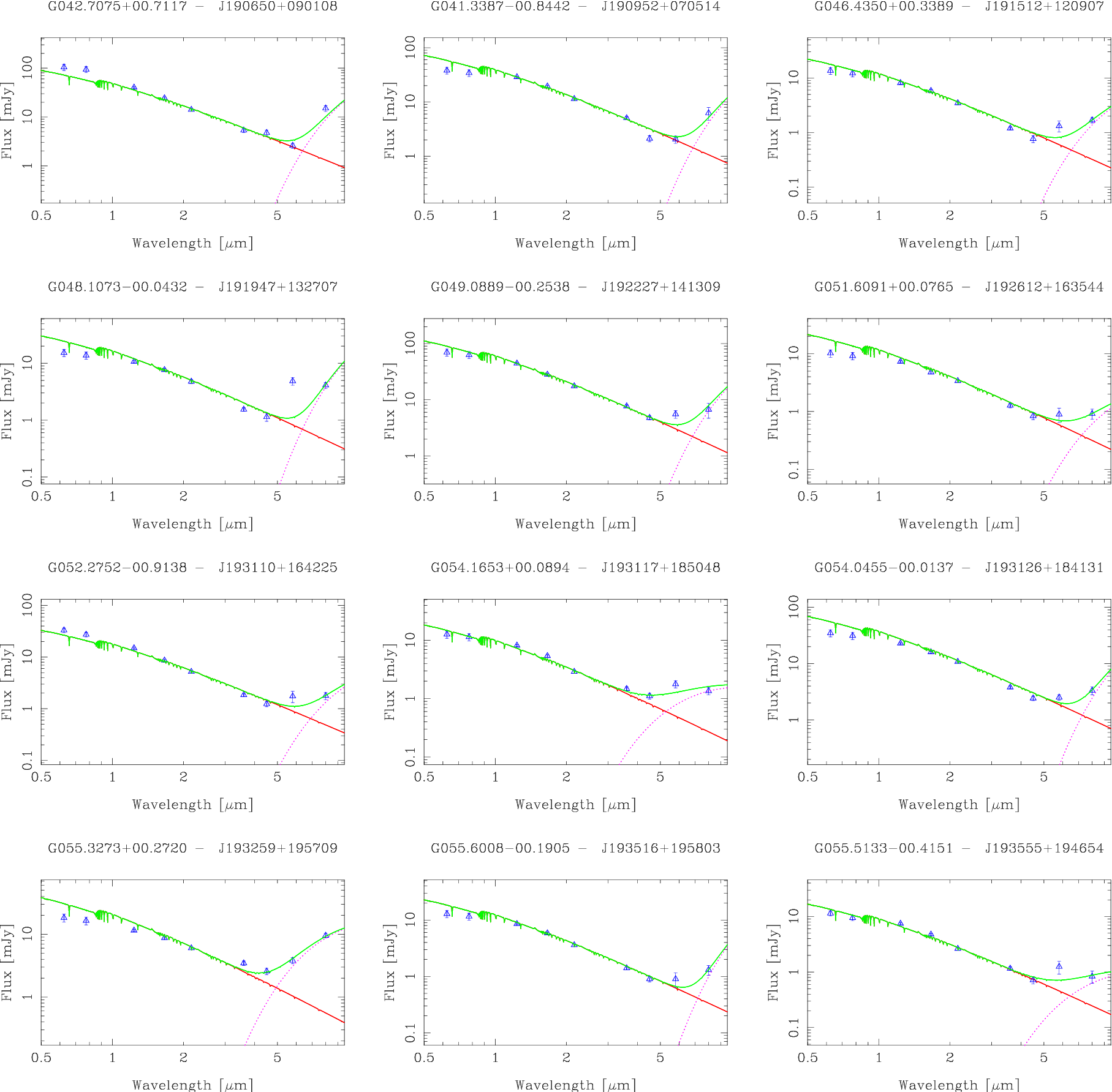}
\caption{ Dereddened optical IPHAS, 2MASS and GLIMPSE photometry for
the IPHAS-selected A-type stars with 8-$\mu$m excesses (blue triangles
with error bars denoting the 1$\sigma$ photometric error of each
measurement). For each star, the red solid line corresponds to the
reference photospheric SED normalised to the K-band flux. The pink
dotted line represents the black-body that, added to the stellar SED,
minimizes the $\chi^{2}$ fit to the data (solid green line). Only the
2MASS and GLIMPSE data points were included in the best-fitting
routine.\label{fig:sed1}}
\end{center}
\end{figure}

\begin{figure}
\begin{center}
\includegraphics[scale=0.5,angle=0,clip=true]
{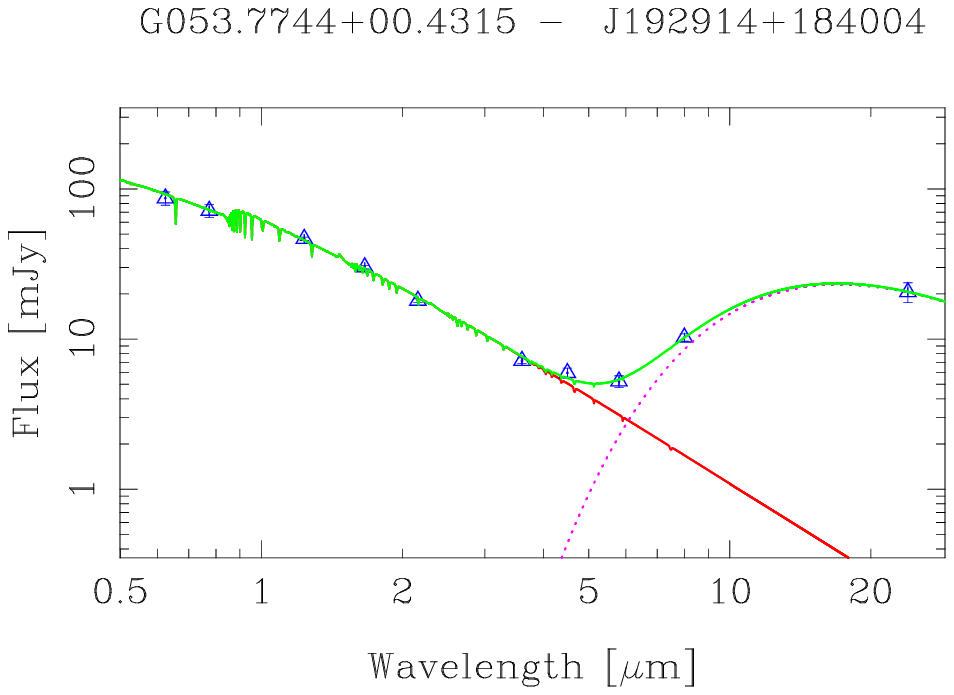}
\caption{ Dereddened optical IPHAS, 2MASS and GLIMPSE photometry for
the IPHAS-selected A-type stars with excesses at 8~$\mu$m and
24~$\mu$m (blue triangles with error bars denoting the 1$\sigma$
photometric error of each measurement). For each star, the red solid
line corresponds to the reference photospheric SED normalised to the
K-band flux. The pink dotted line represents the black-body that,
added to the stellar SED, minimizes the $\chi^{2}$ fit to the data
(solid green line). Only the 2MASS and GLIMPSE data points were
included in the best-fitting routine.\label{fig:sed2}}
\end{center}
\end{figure}

%
%
%
%


\begin{figure}
\begin{center}
\includegraphics[scale=0.8,angle=0,clip=true]
{f9.eps}
\caption{{\it{Left-hand panel}}: Distributions of observed and
 dereddened r$^{\prime}$ and [8] magnitudes for the 2692 A-type
 stars. After dereddening, good agreement between the IPHAS and
 Spitzer photometry is found, in accord with most stars being A-types.
 {\it{Right-hand panel}}: Histogram showing the distance distribution
 of the GLIPHAS sample, together with the IRAC 8-$\mu$m and MIPS
 24-$\mu$m limiting distances for an unreddened A3V star. The dotted
 lines represent the limiting distances for detection of a 10~mJy and
 1~mJy A3V star respectively, while the dashed line shows the 1~mJy MIPS
 limit. The limiting distances for unreddened A-type dwarfs in the IPHAS 
 and 2MASS surveys fall well beyond the
 range of this plot ($>$20~kpc and $>$4~kpc, respectively).
\label{histomagd}}
\end{center}
\end{figure}

\begin{figure}[h!]
\begin{center}
\includegraphics[scale=0.7,angle=0,clip=true]
{f10.eps}
\caption{ Distributions of best-fit blackbody dust temperatures
({\it{left-hand panel}}) and fractional infrared excesses (L$_{\rm
IR}$/L$_{\star}$, {\it{right-hand panel}}) for sources with
excesses at two or more wavelengths.
\label{histos}}
\end{center}
\end{figure}



%

\end{document}